\documentclass[trans]{IEEEtran}
\usepackage{hyperref}
\usepackage{lettrine}
\usepackage{epsfig}
\usepackage{footnote}
\usepackage{cite}
\usepackage{float}
\usepackage{amssymb}
\usepackage{amsthm}
\usepackage{amsmath,mathabx}
\usepackage{color}
\usepackage{amsfonts}
\usepackage{epstopdf}
\usepackage{tabularx,tabulary}
\usepackage[font=footnotesize]{caption}
\usepackage{graphicx}
\usepackage{caption}
\usepackage{subcaption}
\usepackage{wrapfig}
\usepackage{dblfloatfix}    % To enable figures at the bottom of page
\usepackage{bbm}
\usepackage{algorithm}
\usepackage{algpseudocode}

\makeatletter
\algrenewcommand\ALG@beginalgorithmic{\small}
\makeatother

\makeatletter
\def\BState{\State\hskip-\ALG@thistlm}
\makeatother
\usepackage{bbold}

\newcommand{\vect}[1]{\boldsymbol{#1}}

\DeclareMathOperator{\argmin}{argmin}
\DeclareMathOperator{\argmax}{argmax}

\DeclareMathOperator{\erfc}{erfc}

\begin{document}
\title{Opportunistic Routing for Opto-Acoustic\\ Internet of Underwater Things}
\author{
Abdulkadir~Celik,~\IEEEmembership{Senior Member,~IEEE,} Nasir~Saeed,~\IEEEmembership{Senior Member,~IEEE,} \\
 Basem Shihada,~\IEEEmembership{Senior Member,~IEEE,} Tareq Y. Al-Naffouri,~\IEEEmembership{Senior Member,~IEEE,} \\
 and Mohamed-Slim~Alouini,~\IEEEmembership{Fellow,~IEEE}

\thanks{Authors are with Computer, Electrical, and Mathematical Sciences and Engineering Division at King Abdullah University of Science and Technology (KAUST), Thuwal,  23955-6900, KSA (Corresponding Author: Abdulkadir Celik). A part of this work was presented in IEEE WCNC 2019 in Marrakesh, Morocco \cite{Celik2018Sector}.}
}

%\markboth{Submitted to IEEE Internet of Things Journal}{Special Issue on
%Internet of Things for Smart Ocean}

\maketitle

\begin{abstract}
Internet of underwater things (IoUT) is a technological revolution that could mark a new era for scientific, industrial, and military underwater applications. To mitigate the hostile underwater channel characteristics, this paper hybridizes underwater acoustic and optical wireless communications to achieve an ubiquitous control and high-speed low-latency networking performance, respectively. Since underwater optical wireless communications (UOWC) suffers from limited range, it requires effective multi-hop routing solutions. In this regard, we propose a \underline{Sect}or-based \underline{O}pportunistic \underline{R}outing (SectOR) protocol. Unlike the traditional routing (TR) techniques which unicast packets to a unique relay, opportunistic routing (OR) targets a set of candidate relays by leveraging the broadcast nature of the UOWC channel. OR improves the packet delivery ratio as the likelihood of having at least one successful packet reception is much higher than that in conventional unicast routing. Contingent upon the performance characterization of a single-hop link, we obtain a variety of local and global metrics to evaluate the fitness of a candidate set (CS) and prioritize the members of a CS. Since rate$\leftrightarrow$error and range$\leftrightarrow$beamwidth tradeoffs yield different candidate set diversities, we develop a candidate filtering and searching algorithm to find the optimal sector shaped coverage region by scanning the feasible search space. Moreover, a hybrid acoustic/optic coordination mechanism is considered to avoid duplicate transmission of the relays.  Numerical results show that SectOR protocol can perform even better than an optimal unicast routing protocol in well-connected UOWNs.

 \end{abstract}
\IEEEpeerreviewmaketitle

%\begin{IEEEkeywords} 
%Hybrid Acoustic-Optic Networks, Visible Light Communications, Optical Wireless Communications, Opportunistic Routing, internet of things. 
%\end{IEEEkeywords}

\section{Introduction}
\label{sec:intro}

\lettrine{O}{ceans} cover approximately \%71 of the Earth's surface and provide significant benefits to humanity, such as climate regulation, transportation, food supply, recreation, medicine, and a variety of natural resources \cite{Celik2018Survey}. In order to protect oceans and reap their full profits at the same time, it is crucial to transport, store, organize and process the surging amount of data acquired from underwater sensors and fixed/mobile maritime terminals. To this aim, the internet of underwater things (IoUT) is a technological revolution towards integrating physical and digital worlds by interconnecting smart underwater objects \cite{celik2019software}. Thus, IoUT could mark a new era for scientific, industrial, and military underwater applications, e.g., disaster prevention, offshore exploration, environmental monitoring, tactical surveillance, and assisted navigation.

Emerging IoUT applications demand an ambitious quality of service, which necessitates high-speed, ultra-reliable, and low latency underwater networking solutions. However, such goals pose daunting challenges for most electromagnetic frequencies due to the hostile channel impediments of the aquatic medium. Although acoustic communication is a proven technology that is widely adopted by existing underwater applications, its limited bandwidth and low achievable rates are not sufficient for emerging IoUT applications. In spite of its desirable omnidirectional transmission and long communication range, the low propagation speed of acoustic waves (1500 m/s) yields a high latency that disrupts the proper functioning of long-range applications, especially for real-time operation and synchronization tasks \cite{Melike2011Survey}.

Recently, underwater optical wireless communication (UOWC) has gained attention by its advantages of higher bandwidth, lower latency, and enhanced security. Nonetheless, UOWC systems are mainly restrained by its directivity and short communication range. The following phenomena primarily drive these restrictions \cite{Celik2018Survey}: The transmitted light intensity reduces along the propagation path, this energy dissipation is referred to as absorption and caused by the transformation of photon energy into the internal energy of the absorber (e.g., heat, chemical, etc.). Unlike the ballistic photons, some other photons deflect from the default propagation path; this is also known as scattering and caused either by water particles of size comparable to the carrier wavelength (i.e., diffraction) or by constituents with different refraction index (i.e., refraction). Therefore, the relation between absorption and scattering primarily characterizes the fundamental tradeoff between range and beam divergence angle. 

Therefore, multihop UOWC is of the utmost importance to extend communication ranges and realize underwater optical wireless networks (UOWNs) in real-life. In particular, the design and provision of advanced routing protocols top the list of open networking problems as it couples medium access control issues with unique physical layer characteristics of UOWCs. First and foremost, existing routing protocols developed for omnidirectional terrestrial wireless sensor networks and underwater acoustic networks cannot be used for UOWNs in a plug-and-play fashion. Due to the directed nature of the light propagation, the coverage region of a light source is in a sector shape whose central angle (i.e., the divergence angle of the light beam) and radius (i.e., communication range) are inversely proportional. Hence,  a wide divergence angle (e.g., light-emitting diodes) allows reaching nearby neighbors, whereas employing a narrow divergence angle (e.g., lasers) renders communicating with a distant node \cite{Celik2018Modeling}. While the latter requires less number of hops to reach the destination at the cost of equipping the transceivers with accurate pointing-acquisitioning-tracking (PAT) mechanisms, the former may operate without PAT at the expense of a higher number of hops and power consumption \cite{Celik2019Endtoend}. 

\begin{figure}[t]
\begin{center}
\includegraphics[width=0.99\columnwidth]{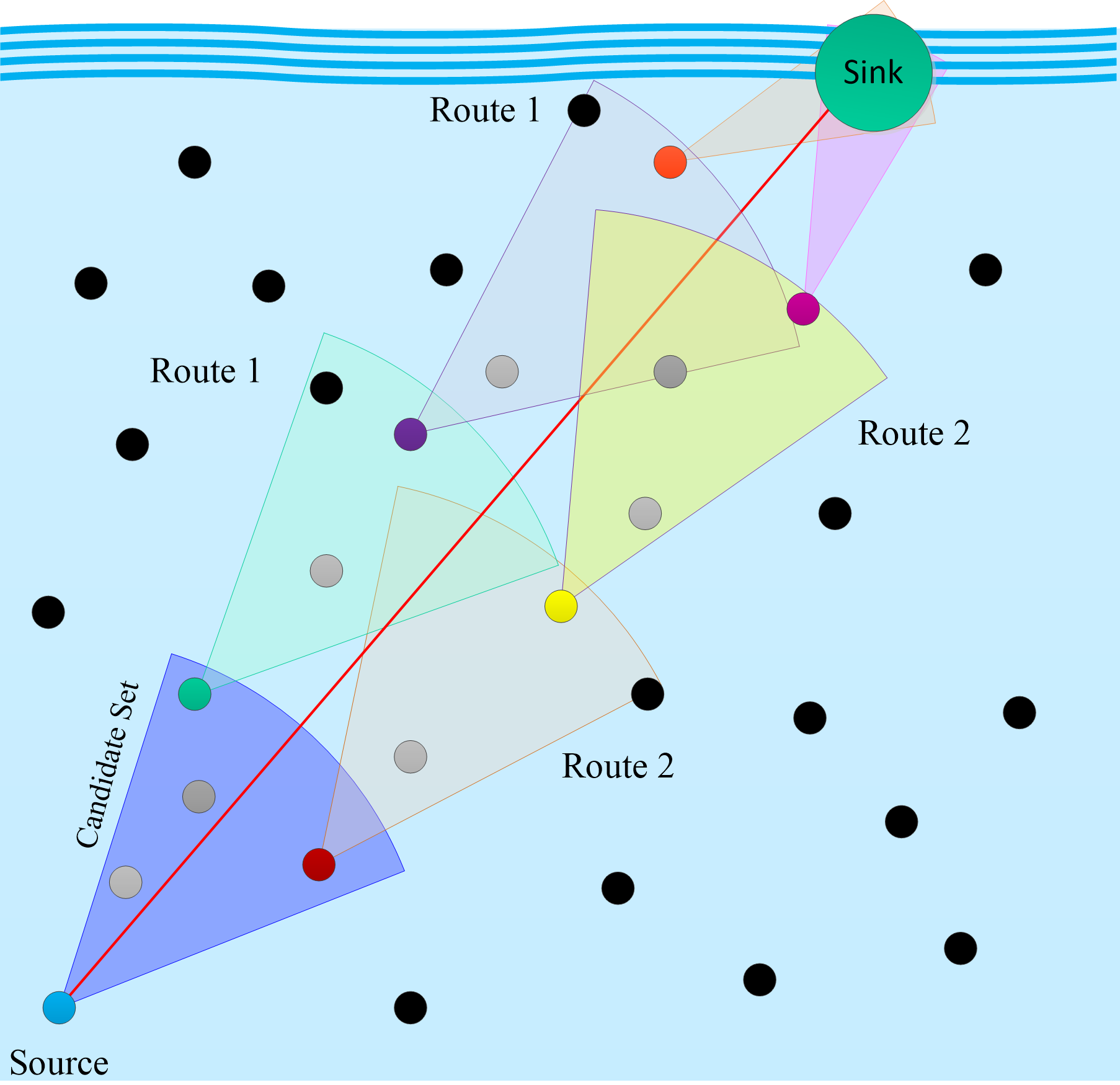}
\caption{Illustration of UOAN and SectOR protocol.}
\label{fig:uown}
\end{center}
\end{figure}

Apart from the traditional unicast routing (TUR) protocols that transmit packets to a unique next-hop forwarder, opportunistic routing (OR) broadcasts packets to a set of candidate nodes. TUR protocols retransmit unicast lost packets to the forwarder, which are eventually discarded after a specific number of re-transmission. On the contrary, by leveraging the broadcast nature of UOWC, OR involves other candidates in forwarding the packets if the chosen forwarder fails to receive the packet.  For instance, Fig. \ref{fig:uown} demonstrates two different routes: The former is the route when the highest priority node (green) successfully receives the packet while the latter is over the second-highest priority node (red) when the highest priority fails to receive packet correctly. Hence, OR improves the packet delivery ratio as the likelihood of having at least one successful packet reception is much higher than that in TUR. 

In this way, OR provides more energy, throupput, and delay efficient communications by reducing the expected number of re-transmissions. Furthermore, OR is especially suitable to UOWNs because of the connection interruptions caused either by underwater channel impediments (e.g., pointing errors, misalignment, turbulence, etc.) or sea creatures passing through the transceivers' line-of-sight. Nonetheless, OR requires effective cooperation and coordination mechanisms among the candidate nodes to avoid duplicate transmissions and collisions. To this end,  this paper proposes a distributed \textbf{\underline{Sect}}or-based \textbf{\underline{O}}pportunistic \textbf{\underline{R}}outing (SectOR) protocol. Being inspired by the sector-shaped coverage region of the light sources, the SectOR discovers routing path towards the destination by exploiting local or global OR metrics based on topology information available at IoUT nodes.

\subsection{Related Works}
The IoUT concept is surveyed in \cite{DOMINGO2012IouT} and \cite{Kao2017IoUT} where IouT and its potential applications are presented in a similar fashion to underwater acoustic sensors networks. A more comprehensive approach is provided in \cite{AKYILDIZ20161} where authors consider software-defined IoUT nodes that can employ acoustic, optic, and magnetic induction signals to overcome the peculiarities of the underwater environment. A software-defined opto-acoustic network architecture design is also proposed in \cite{celik2019software} where authors explain inextricably interwoven relations among functionalities of different layers and introduce network function virtualization (NFV) to realize application specific cross-layer protocol suites through an NFV management and orchestration system.   

Although physical layer issues of UOWC is relatively mature, its networking aspects still stays unexplored. Recent efforts on UOWNs can be exemplified as follows: Assuming a Poisson point process based spatial distribution, Saeed et. al. analyzed the $k$-connectivity of UOWNs  \cite{Celik2018Connectivity}. In \cite{Jamali2016performance}, authors characterized the performance of relay-assisted underwater optical CDMA system where multihop communication is realized by chip detect-and-forward method. Similarly, Jamali et. al. consider the performance anaylsis of multihop UOWC using  decode-and-forward (DF) relaying \cite{Jamali2017multihop}. In \cite{Celik2018Modeling}, we addressed modeling and end-to-end performance analysis of multihop UOWNs under both DF and amplify-and-forward methods. We extended this work in \cite{Celik2019Endtoend} to investigate the impacts of location uncertainty on key performance metrics such as achievable rate, bit error rate, and power consumption. Excluding \cite{Celik2019Endtoend,Celik2018Modeling}, these works do not deal with the effective UOWN routing protocols which is of utmost importance to extend the limited communication range of UOWCs. In \cite{Celik2019Endtoend,Celik2018Modeling}, proposed protocols follow a traditional unicast routing approach which adapts shortest path algorithms to find paths with minimum distance/error/power consumption or maximum rate. OR has been extensively studied for terrestrial wireless networks \cite[and references therein]{Chakchouk2015SurveyOR}. OR is also applied to underwater acoustic networks \cite{Menon2016Comparative,Darehshoorzadeh2015Underwater,Coutinho2016Geographic,Coutinho2014GEDAR}. To the best of authors knowledge, this work is first to develop an OR protocol for UOWNs.

\subsection{Main Contributions}
Our main contributions in this paper can be summarized as follows:
\begin{itemize}
\item  
Based on unicast link performance analysis, the performance of broadcast links are characterized in terms of data rate, maximum range, packet delivery ratio, and expected number of retransmissions. Using these performance characterizations as building blocks, we then developed both local and global OR metrics such as distance progress, energy efficiency, and low latency. 

\item 
Since pointing direction and range$\leftrightarrow$beamwidth tradeoff yield different candidate set diversities, a candidate filtering and selection technique is proposed to find a pointing and divergence angle. By manipulating the pointing angles and leveraging adaptive beamwidths, we find the candidate set that delivers the best OR metric of interest. Based on this technique, each node maintains its best pointing and divergence angle, and forward received packets along with the priority order of its candidates. Moreover, candidate coordination is facilitated by acoustic communications to mitigate the directivity and range limitations of UOWC.
\end{itemize}

\subsection{Notations and Paper Organization}
Throughout the paper, sets and their cardinality are denoted with calligraphic and regular uppercase letters (e.g., $| \mathcal{X}|=X$), respectively. Vectors and matrices are represented in lowercase and uppercase boldfaces (e.g., $\vect{x}$ and $\vect{X}$), respectively. Superscripts $s$, $d$, $i$, and $j$ are used for indexing source, destination, current forwarder, and next forwarder nodes, respectively. The optimal/best values of variables are always marked with superscript $\star$, e.g., $x_{sd}^{i\star}$. 
%For a table of symbols, we refer interested readers to Table \ref{tab:symbols} in the Appendix.   

The remainder of the paper is organized as follows: Section \ref{sec:sysmod} introduces the network and channel models. Section \ref{sec:performance} analysis the performance of unicast and broadcast links. These performance characterizations are then used to develop local and global OR metrics in Section \ref{sec:OR_Metrics}. Section \ref{sec:SectOR} provide the details of the proposed SectOR protocol and summarizes the algorithmic implementation. Section \ref{sec:res} presents the numerical results. Finally, Section \ref{sec:conc} concludes the paper with a few remarks. 

\section{Underwater Opto-Acoustic Networks}
\label{sec:sysmod}
In this section, we introduce the UOAN of interest, present the UOWC channel model, and explain the tradeoff between communication range and beamwidth. 
\subsection{Network Model}
\label{sec:netmod}
We consider a UOAN  that consists of a single sink/surface station and $\rm{M}$ IoUT nodes, as demonstrated in Fig. \ref{fig:uown}. IoUT nodes are equipped with low-cost optical transceivers to enable UOWC in both forward and backward directions. Light sources are assumed to be capable of adapting their beamwidth and communication range by adjusting the divergence angle \cite{LoPresti06}. Although optical transceivers are primarily employed to deliver a large volume of sensing data via high-speed UOWC links, the limited range and directivity of UOWC hinder its ability to serve as a reliable control medium for network management tasks. Thanks to its omnidirectional propagation characteristics, each node also has a single acoustic transceiver to provide the network with highly connected control links. The sink station is responsible for aggregating data from sensors and disseminating this information to mobile or onshore sinks. Since IoUT data is generally useful only if it is geographically tagged to an accurate sensing location, we assume that each node is aware of its own location ($\boldsymbol{\ell}_i, i \in [1,\rm{M}]$) along with the neighbors within its acoustic communication range.  Underwater location information can be obtained either by a fully optic \cite{Celik2019Localization, Celik2019Performance} or hybrid opto-acoustic \cite{Celik2017HybAcoOpt} network localization methods. 
%By default, all sensors keep the orientation of their body frames and pointing vector of the optical transmitter to be directed toward the sink node. 

%Referring to $\epsilon$ as the error of underlying localization algorithm, we denote the actual and estimated locations of node $i \in [1,\rm{M}]$ ($n_i$) by $\boldsymbol{\ell}_i=[x_i, y_i]$ and $\boldsymbol{\tilde{\ell}}_i=[x_i+\epsilon, y_i+\epsilon]$, respectively.  
%While seabed sensors are fixed to the ground, others are either moored or buoyed as shown in Fig. \ref{fig:uown} where red-colored buoyed nodes closer to the surface are required to relay all the uplink information to the sink. 
%Spatial distribution of nodes is assumed to follow a homogeneous Poisson point process of density $\mu$. 

\begin{figure}[!t]
\begin{center}
\includegraphics[width=1\columnwidth]{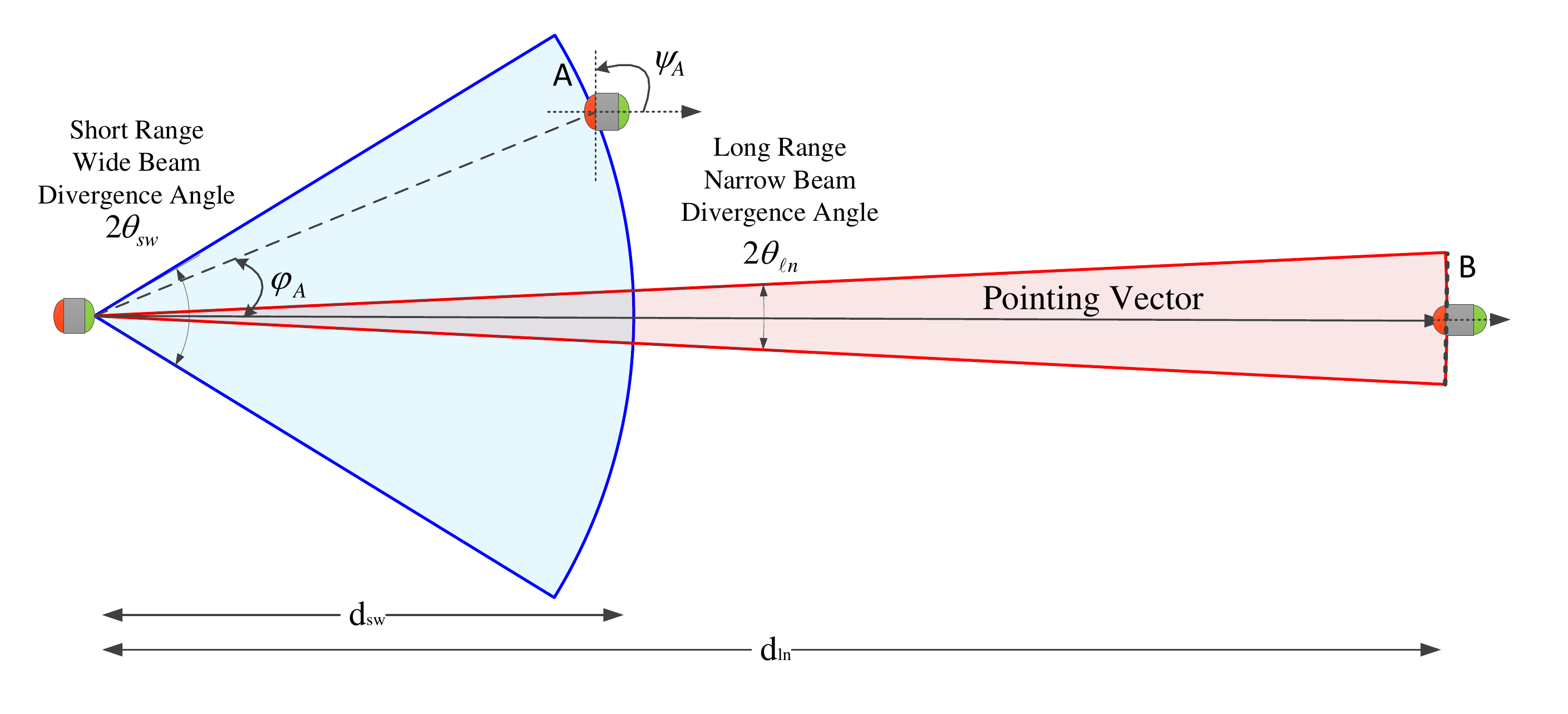}
\caption{Illustration of a single-hop link and the tradeoff between divergence angle and communication range.}
\vspace{-0.7cm}
\label{fig:sys}
\end{center}
\end{figure}

\subsection{Channel Model}
\label{sec:channel}

%\subsubsection{Optical Channels}
%\label{sec:optic}
According to the Beer's law, absorption and scattering effects of the aquatic medium can be characterized by an extinction coefficient $c(\lambda)= a(\lambda)+b(\lambda)$ where $\lambda$, $a(\lambda)$, and $b(\lambda)$ denote the carrier wavelength, absorption coefficient, and scattering coefficient, respectively. Based on Beer-Lambert formula, the propagation loss factor between two generic IoUT nodes $n_i$ and $n_j$ is defined as follows 
\begin{equation}
\label{eq:BL}
\mathrm{BL_i^j}=\exp \left \{ -c(\lambda) \frac{d_i^j}{\cos(\varphi_i^j)} \right\},
\end{equation}
where $d_i^j$ is the perpendicular distance between the nodes and $\phi_i^j$ is the angle between the receiver plane and the transmitter-receiver trajectory, as shown in Fig. \ref{fig:sys} where $n_j$ is located at point $A$. In case of a perfect alignment, \eqref{eq:BL} reduces to $\mathrm{BL_i^j}=\exp \left \{ -c(\lambda)d_i^j \right\}$ if $n_j$ is located at point $B$. On the other side, the geometrical loss is a result of spreading the light beam to an area larger than the receiver aperture size $A_j$ and can be given for a semi-collimated transmitter emitting a Gaussian beam by
\begin{equation}
\label{eq:GL}
\mathrm{GL_i^j}= \left(\frac{A_{j} \cos(\varphi_i^j) }{\theta_{1/e}^i d_i^j}\right)^2
\end{equation}
where $\theta_{1/e}^i $ is the is full-width beam divergence angle of $n_i$ that is measured at the point where the light intensity drops to $1/e$ of its peak. In the case of perfect alignment, \eqref{eq:GL} reduces to the approximation given in \cite{Poliak2012link}. Accordingly, the channel gain between $n_i$ and $n_j$ is given by the product of \eqref{eq:BL} and \eqref{eq:GL} as  
\begin{equation}
\label{eq:PL}
\mathrm{G_i^j}= \left(\frac{A_{j} \cos(\varphi_i^j) }{\theta_{1/e}^i d_i^j}\right)^2 \exp \left \{ -c(\lambda) \frac{d_i^j}{\cos(\varphi_i^j)} \right\},
\end{equation}
that is merely based on the received ballistic photons which propagate without being disturbed by the scattering effects. That is, \eqref{eq:PL} neglects all of the scattered photons received by $n_j$ by assuming their total loss. By modifying \cite[Eq. (4)]{Elamassie2018performance}, scattered rays can be taken into account as follows  
\begin{equation}
\label{eq:PLmodf}
\mathrm{G_i^j}=\left(\frac{A_{j} \cos(\varphi_i^j) }{\theta_{1/e}^i d_i^j}\right)^2 \exp \left \{ - \frac{c(\lambda)d_i^j}{\cos(\varphi_i^j)}\left(\frac{A_{j} \cos(\varphi_i^j) }{\theta_{1/e}^i d_i^j}\right)^\alpha \right\}
\end{equation}
where $\alpha$ is a correction coefficient which can be determined based on parameters such as $c(\lambda)$, $A_j$, $\theta_{1/e}^i$, field-of-view (FoV) angle of the receiver, etc. By analyzing \eqref{eq:PL} and \eqref{eq:PLmodf},  one can gain valuable insight into the tradeoff between divergence angle and communication range. As illustrated in Fig. \ref{fig:sys}, a wide divergence angle results in a short communication range so that the source can reach the neighbor nodes within its proximity. On the other hand, a narrow divergence angle helps to reach a distant receiver, which naturally requires an agile and accurate PAT mechanism to sustain a reliable communication link. 
%\subsubsection{Acoustic Channels}
%\label{sec:optic}

\section{Performance Characterization
of\\Single-Hop Unicast Links}
\label{sec:performance}

In this section, we characterize the performance of unicast and broadcast links which are used as building blocks of the OR metrics developed in the next section. 

\subsection{Unicast Links in TUR}
\label{sec:unicast}
The source node $n_s$ groups messages destined to the destination node $n_d$ into packets, each has a length of $L$ bits that consist of a header and a payload. While control messages (e.g., destination address, next forwarder, ACK signaling, etc.) are included in the header, data is encapsulated in the payload that is extracted and used by IoUT applications. TUR paths are formed by consecutive unicast links, i.e., data is forwarded to a unique node at each hop. Thus, we first characterize the performance of a unicast link in terms of distance, reliability, and achievable rates. 

Let us consider an arbitrary multihop path between $n_s$ and $n_d$, $s \rightsquigarrow d=\{s,\ldots,i,j,\ldots,d\}$. Assuming that the number of photons follows a Poisson Process, photon arrival rate from $n_i$ to $n_j$ is given by \cite{Arnon10underwater} 
\begin{equation}
\label{eq:photon_count}
f_{ij}=\frac{P_{rx}^j \eta_c^j \lambda}{R_{i}^j T \hslash c},
\end{equation}
where $P_{rx}^j=P_{tx}^i \eta_{tx}^i \eta_{rx}^j \mathrm{G_i^j}$ is the received power by $n_j$, $P_{tx}^i$ is the transmission power of $n_i$, $\eta_{tx}^i$ ($\eta_{rx}^j$) is the transmitter (receiver) efficiency of $n_i$ ($n_j$), $\eta_c^j$ is the detector counting efficiency of $n_j$, $\rm{R}_{i}^j $ is the data rate, $T$ is pulse duration, $\hslash$ is Planck's constant, and $c$ is the underwater speed of light. As per the central limit theorem, the Poisson distribution can be approximated by a Gaussian distribution if the number of received photons is large enough.
For intensity-modulation/direct-detection (IM/DD) with on-off keying (OOK) modulation, bit error rate (BER) of the link between $n_i$ and $n_j$ is given by \cite{Celik2019Endtoend}
\begin{equation}
\label{eq:BER}
\overrightarrow{\rm{BER}}_i^j=\frac{1}{2}\erfc \left(\sqrt{\frac{T}{2}} \left( \sqrt{f_{ij}^1} - \sqrt{f_{ij}^0}\right) \right)
\end{equation}
where $\erfc(\cdot)$ is the complementary error function, $f_{ij}^1=f_{ij}+f_{dc}+f_{bg}$  and $f_{ij}^0=f_{dc}+f_{bg}$are the numbers of photon arrivals when binary 1 and binary 0 are transmitted, respectively, $f_{dc}$ is the additive noise due to dark counts, and $f_{bg}$ is the background illumination noise. Accordingly, the packet error rate (PER) and packet delivery ratio (PDR) can be given by 
\begin{align}
\label{eq:PER}
\overrightarrow{\rm{PER}}_i^j&=1-\left(1-\overrightarrow{\rm{BER}}_i^j\right)^L \text{ and} \\
\overrightarrow{\rm{PDR}}_i^j&=1-\overrightarrow{\rm{PER}}_i^j=\left(1-\overrightarrow{\rm{BER}}_i^j\right)^L,
\end{align}
respectively. 

For a given PER, data rate between $n_i$ and $n_j$ is then derived by using \eqref{eq:photon_count}-\eqref{eq:PER} as
\begin{equation}
\label{eq:rate}
\rm{R}_i^j=\frac{\eta_j^c \lambda}{2 \hslash c } \left[ \frac{\sqrt{P_{rx}^j+P_{dc}+ P_{bg}}-\sqrt{P_{dc}+ P_{bg} }}{\erfc^{-1} \left(2 a\right) }\right]^2
\end{equation}
where $P_{dc}$ is the dark count noise power, $P_{bg}$ is the background noise power, and $a=1- \left(1-\overrightarrow{\rm{PER}}_i^j\right)^{\frac{1}{L}}$. For a given data rate $\rm{R}_i^j$, the communication range between $n_i$ and $n_j$ is obtained from the perpendicular distance by using \eqref{eq:PL}-\eqref{eq:BER} as follows
\begin{equation}
\label{eq:dist}
\rm{D}_i^j= \frac{1}{\cos(\varphi_i^j)}\left(\frac{2}{(\alpha-1)b_3} W_0 \left[\frac{(\alpha-1)\left(b_1b_2\right)^{\frac{\alpha-1}{2}}b_3}{2} \right] \right)^{\frac{1}{1-\alpha}},
\end{equation}
where $W_0(\cdot)$ is the principal branch of product logarithm, $$b_1=\frac{\left[ \sqrt{\frac{2 \rm{R}_i^j \hslash c }{\eta_j^c \lambda}} \erfc^{-1}(2a)+\sqrt{P_{dc}+P_{bg}} \right]^2-P_{dc}-P_{bg}}{P_{tx}^i \eta_{tx}^i \eta_{rx}^j },$$ $$b_2=\left( \frac{\theta_{1/e}^i}{A_j \cos(\phi_i^j)} \right)^2, \: \text{and } b_3= -\frac{c(\lambda)}{\cos(\varphi_i^j)} 
\left(\frac{A_j\cos(\varphi_i^j)}{\theta_{1/e}^i}\right)^\alpha.$$ 

At this point, it is important to relate the previously discussed range$\leftrightarrow$beamwidth tradeoff with the rate$\leftrightarrow$reliability one. While the former is specific to OWC since link distance reduces as the divergence angle increases, the latter is common for any communication systems as data rate and PDR are inversely proportional to each other, which follows from \eqref{eq:rate}.  Following from \eqref{eq:dist}, these two tradeoffs are also coupled as the range is a function of rate, PDR, and divergence angle.

Assuming that the packet is dropped after $\rm{K}$ retransmission attempts, probability of having a successful transmission to $n_j$ in $k$ delivery attempts is given by
\begin{align}
\mathcal{P}_{i}^j(k)&=\left(\overrightarrow{\rm{PER}}_i^j\right)^{k-1} \overrightarrow{\rm{PDR}}_i^j, \: k\in [1,\rm{N}]
\end{align}
Hence, expected number of transmissions (ExNT) can be obtained as
\begin{align}
\label{eq:ExNT}
\rm{N}_i^j&=\sum_{k=1}^{\rm{K}} k \mathcal{P}_{i}^j(k)+ \rm{K} \left(\overrightarrow{\rm{PER}}_i^j\right)^{\rm{K
}}
%&=\left(\overrightarrow{\rm{PDR}}_i^j \right)^{-1}
\end{align}
where $\left(\overrightarrow{\rm{PER}}_i^j\right)^{\rm{N}}$ is the probability of dropping the package. If the ExNT is normalized to the probability of having a successful transmission within $\rm{N}$ retransmission, we obtain unicast ExNT as 
\begin{align}
\label{eq:ExNT_norm}
 \overline{\rm{N}}_i^j&=\frac{\overrightarrow{\rm{N}}_i^j}{\sum_{k=1}^{\rm{K}}\mathcal{P}_{i}^j(n)} = \frac{1}{\overrightarrow{\rm{PDR}}_i^j},
\end{align}
which is independent of $\rm{N}$.

\begin{figure}[t]
\begin{center}
\includegraphics[width=0.75\columnwidth]{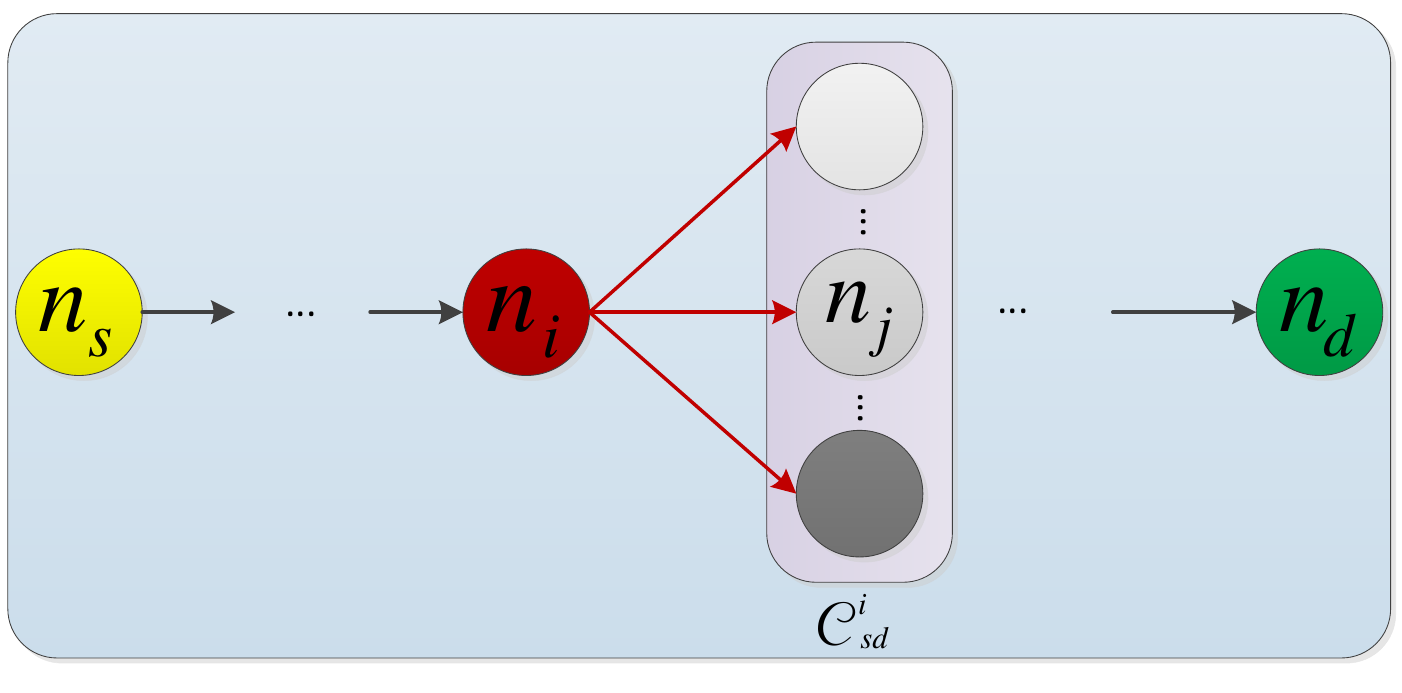}
\caption{Illustration of opportunistic routing and broadcast links.}
\label{fig:OR}
\end{center}
\end{figure}

\subsection{Broadcast Links in OR}
\label{sec:broadcast}

Unlike the TUR protocols, OR selects a candidate set (CS) that can overhear the broadcast packets and forward them to the next-hop in a prioritized and coordinated manner. Assuming that $n_i$ is one of the forwarder nodes from $n_s$ to $n_d$ [c.f. Fig. \ref{fig:OR}], $\mathcal{C}_{sd}^i$ is denoted as the candidate index set of the forwarder node $n_i$\footnote{ $\mathcal{C}_{sd}^i$ may also include the destination node if it is within the communication range.}. For simplicity, we assume that $\mathcal{C}_{sd}^i$ is ordered in the descending order of forwarding priority. That is, the $k^{th}$ member of $\mathcal{C}_{sd}^{i}$ attempts to forward packet only if the $j^{th}$ member fails, $k>j$. In this case, packet delivery fails if none of the CS members successfully receives the packet, i.e., 
\begin{equation}
\label{eq:OR_PDR}
\rm{PER}_{sd}^{i}=\prod_{j \in \mathcal{C}_{sd}^i} \overrightarrow{\rm{PER}}_i^j.
\end{equation}
% Accordingly, the opportunistic PDR can be given by 
% \begin{equation}
% \label{eq:opp_PDR}
% \rm{PDR}_{sd}^{i}=1-\rm{PER}_{sd}^{i}=1-\prod_{j \in \mathcal{C}_{sd}^i} \left[ 1-\left(1-\overrightarrow{\rm{BER}}_i^j\right)^L \right]
% \end{equation}
% which is the probability that at least one of the candidates successfully receives the packet sent by $n_i$. 
Accordingly, successful packet forwarding ratio is obtained as 
\begin{equation}
\label{eq:OR_PFP}
\rm{SFR}_{sd}^{ij}= \overrightarrow{\rm{PDR}}_i^j \prod_{k=1}^{j-1}  \overrightarrow{\rm{PER}}_i^k
\end{equation}
which is the probability that $n_j$ successfully receives the packet from $n_i$ given that higher priority candidates $\mathcal{C}_{sd}^{ik}$ , $k<j$, fail. 
Hence, ExNT in the OR scheme is given by 
\begin{equation}
\label{eq:ExNT_OR}
\rm{N}_{sd}^i= \sum_{k=1}^{\rm{K}} k \left(\rm{PER}_{sd}^{i}\right)^{k-1} \rm{PDR}_{sd}^{i}+ \rm{K} \left(\rm{PER}_{sd}^{i}\right)^{\rm{K}}
\end{equation}
where the first term is the ExNT for successfully delivering a package to $n_j$ and the second term accounts for the package drop event. As in \eqref{eq:ExNT_norm}, the ExNT normalized to the success probability is given by 
\begin{equation}
\label{eq:ExNT_OR_norm}
\overline{\rm{N}}_{sd}^i=\frac{1}{\rm{PDR}_{sd}^{i}}=\frac{1}{1-\prod_{j \in \mathcal{C}_{sd}^i} \overrightarrow{\rm{PER}}_i^j}.
\end{equation} 
which is referred to as broadcast ExNT in the rest of the paper. 

\section{Opportunistic Routing Metrics}
\label{sec:OR_Metrics}

OR metrics play a crucial role in the performance of the routing protocol since it has a direct impact on the candidate selection and prioritization outcomes. Based on the available network state information at each node, the OR metrics can be classified into \textit{local} and \textit{global} metrics, which require information from neighboring nodes and the entire network topology, respectively. We denote $\sphericalangle(d,\theta_{1/e}^i, \psi_i)$ as the sector-shaped coverage region of $n_i$ with divergence angle $\theta_{1/e}^i$ centered at the pointing angle $\psi_i$. Notice that one can alter the sector-shaped optical coverage area of $n_i$ by changing $\psi_i$ and $\theta_{1/e}^i$. That is, the elements of $\mathcal{C}_{sd}^{i} (\psi_i,\theta_{1/e}^i )$ vary with different pairs of $\psi_i$ and $\theta_{1/e}^i$. For the sake of clarity, we omit these parameters from $\mathcal{C}_{sd}^{i} (\psi_i,\theta_{1/e}^i)$, and focus our attention on a single CS, $\mathcal{C}_{sd}^i$, throughout this section. Next, we present local and global OR metrics that account for different routing objectives. 

\subsection{Local Opportunistic Routing (LOR) Metrics}
Local OR metrics are generally preferred to avoid the cost of updating and storing the entire topology state. We assume that each node has the location information of itself, one-hop neighborhood (i.e., acoustic coverage), and the destination (i.e., the sink). In what follows, we introduce local OR metrics for distance progress, energy consumption, and delay.  

\subsubsection{Distance Progress}
One of the most common local OR metrics is the distance progress (DP), which selects and orders the candidate according to their closeness toward the destination \cite{ZorziDP2003}. The DP metric for $n_j \in \mathcal{C}_{sd}^i$is given as 
\begin{equation}
\label{eq:DP}
\rm{DP}_{sd}^{ij}=\left(\Vert \vect{\ell}_s- \vect{\ell}_d \Vert -\Vert \vect{\ell}_j- \vect{\ell}_d \Vert \right), \forall j \in \mathcal{C}_{sd}^i
\end{equation} 
Accordingly, the prioritized $\mathcal{C}_{sd}^{i}$ for DP metric is given by  
\begin{equation}
 \label{eq:DP_priority}
\mathcal{R}(\mathcal{C}_{sd}^{i})=
\{j\vert \rm{DP_{sd}^{ij}} > \rm{DP_{sd}^{il}}, j<l, \forall (j,l) \in  \mathcal{C}_{sd}^{i}\}.
\end{equation}
Lastly, the DP fitness of $\mathcal{C}_{sd}^i$ is given by
\begin{equation}
\label{eq:DP}
\rm{F_{DP}^i} \left[ \mathcal{C}_{sd}^i \right] = \underset{\forall j \in \mathcal{C}_{sd}^i}{\max} \left\{ \rm{DP}_{sd}^{ij}\right\}.
\end{equation} 
Notice that measuring the OR metric in traveled distance implicitly sets the routing objective to minimize the number of hops. In terrestrial WSNs, the DP metric is limited to the scenario where a very far away candidate is selected merely based on its proximity without accounting for the link quality. Due to the short-range and directed propagation of light in the water, the negative consequences of this limitation can be mitigated by restricting the candidate set to the sector-shaped coverage region. A more advanced version of DP is the expected distance progress (EDP) that considers the average DP \cite{Darehshoorzadeh2012distance} by accounting for the link quality. The EDP metric of $n_i$ is given by 
 \begin{equation}
\label{eq:EDP}
\rm{EDP}_{sd}^{ij}= \rm{DP}_{sd}^{ij} \rm{SFR}_{sd}^{ij}, \: \forall j \in \mathcal{C}_{sd}^{i} 
\end{equation}
which accounts for connectivity, link quality, and distance advancement toward the sink at the same time. Accordingly, the j$^{th}$ element of priority set can be iteratively determined as follows
\begin{equation}
 \label{eq:EDP_priority}
\mathcal{R}_{sd}^{ij}=
\begin{cases}  \:\:\:\:\:\:\underset{k \in \mathcal{C}_{sd}^{i} }{\argmax}& \left(\rm{DP}_{sd}^{ik} \rm{PDR}_i^k \right), j=1 \\
\underset{k \in \mathcal{C}_{sd}^{i} -\bigcup_1^{j-1} \mathcal{R}_{sd}^{ij} }{\argmax}& \left(\rm{DP}_{sd}^{ik} \rm{SFR}_{sd}^{ik} \right), j>1
\end{cases},
\end{equation}
where the first element is determined based on highest individual performance of nodes while the remaining nodes are iteratively determined based on the latest form of the priority set. Lastly, the EDP fitness of $\mathcal{C}_{sd}^i$ is given by
 \begin{equation}
\label{eq:EDP}
\rm{F^i_{EDP}}\left[ \mathcal{C}_{sd}^i \right] =\sum_{ j \in \mathcal{C}_{sd}^i} \rm{DP}_{sd}^{ij} \rm{SFR}_{sd}^{ij}.
\end{equation}

%Nonetheless, the DP and EDP do not factor a crucial UOAN shortcoming in, which is limited connectivity. In addition to the hostile underwater environment and limited transmission range of OWC, UOANs are expected to have a sparse node deployment. In this case, the OR with local metrics may not reach to the destination if the path is routed towards a void region. Hence, local OR metrics should also account for the degree of connectivity.

\subsubsection{Energy Efficiency}
Since IoUT nodes operate on limited battery capacity, energy-efficient OR plays a crucial role in UOAN lifetime maximization. Indeed, the consumed energy increases with the number of transmission attempts, each of which costs an energy dissipation as a result of transmission, reception, and coordination. Hence, the energy cost of making $k$ transmission can be formulated as
\begin{equation}
\label{eq:energy}
\rm{E}_i(k)=k \left[T_s\left( P_{tx}^i  + \sum_{j \in \mathcal{C}_{sd}^i} P_l^j \right)+ P_c T_c^i \right],  
\end{equation}
where $T_s=\frac{L}{R_i}$ is the transmission duration, $P_l^j$ is the listening power consumed by decoding and signal processing circuitry, and $P_c$ is the coordination power consumed by candidates during the coordination duration $T_c^i$\footnote{We revisit the components of $T_c$ during the candidate coordination discussion in Section \ref{sec:coordination}.}. Based on \eqref{eq:energy}, the energy efficiency metric (EEM) for $n_j$ is given by
\begin{equation}
\label{eq:EEM}{}
\rm{EEM}_{sd}^{ij}=  E_i (\rm{N}_i^j)
\end{equation}
where $\rm{N}_i^j$ can be obtained by \eqref{eq:ExNT}. Accordingly, the prioritized $\mathcal{C}_{sd}^{i}$ for EEM metric is given by  
\begin{equation}
 \label{eq:DP_priority}
\mathcal{R}(\mathcal{C}_{sd}^{i})=
\{j\vert \rm{EEM_{sd}^{jk}} < \rm{EEM_{sd}^{il}}, j<l, \forall (j,l) \in  \mathcal{C}_{sd}^{i}\}.
\end{equation} 
Thus, EEM fitness of $\mathcal{C}_{sd}^i$ is given by
 \begin{equation}
\label{eq:EDP}
\rm{F^i_{EEM}}\left[ \mathcal{C}_{sd}^i \right] = E_i( \overline{\rm{N}}_i^j).
\end{equation}
where $ \overline{\rm{N}}_i^j$ is the unicast ExNT given in \eqref{eq:ExNT_norm}.

\subsubsection{Low Latency}
Latency is a critical metric, especially for delay intolerant underwater applications. Similar to EEM, low latency metric (LLM) increases with the number of transmission attempts, each of which costs a delay due to the transmission and coordination among the candidate nodes. Hence, the delay caused by $n$ transmission attempts can be formulated as
\begin{equation}
\label{eq:latency}
\rm{D}_i(n)=n\left[T_s+ T_c^i \right],
\end{equation}
which can be used to calculate the LLM of $n_j$ as follows
\begin{equation}
\label{eq:LLM}
\rm{LLM}_{sd}^{ij}=  D_i(\rm{N}_i^j).
\end{equation}
Similarly, the prioritized $\mathcal{C}_{sd}^{i}$ for LLM metric is given by  
\begin{equation}
 \label{eq:DP_priority}
\mathcal{R}(\mathcal{C}_{sd}^{i})=
\{j\vert \rm{LLM_{sd}^{jk}} < \rm{LLM_{sd}^{il}}, j<l, \forall (j,l) \in  \mathcal{C}_{sd}^{i}\}.
\end{equation} 
Lastly, the LLM fitness of $\mathcal{C}_{sd}^i$ is given by
 \begin{equation}
\label{eq:EDP}
\rm{F^i_{LLM}}\left[ \mathcal{C}_{sd}^i \right] = D_i( \overline{\rm{N}}_i^j).
\end{equation}
where $ \overline{\rm{N}}_i^j$ is the unicast ExNT given in \eqref{eq:ExNT_norm}.

\subsection{Global Opportunistic Routing (GOR) Metrics}
The main objective of the global OR metric is to reduce the ExNT such that end-to-end (E2E) ExNT, energy consumption, and delay is minimized. Since global OR metrics capture the ExNT while taking all possible multipath, they are generally expressed in recursive formulas. Naturally, these require a massive control signaling overhead and computational power. 

For a forwarder node $n_i$, an E2E metric can be obtained by summation of two components: 1) The metric from $n_i$ to its CS $\mathcal{C}_{sd}^i$ and 2) The metric from its candidates to the destination node, $n_d$. In this case, we can rewrite the global version of \eqref{eq:energy} in a recursive form as follows 
\begin{equation}
\label{eq:energy_e2e}
\rm{E}_{i}^{j}(k)=\rm{E}_i(k)+\rm{E}_j(\rm{N}_{sd}^j),
\end{equation}
$\rm{N}_{sd}^j$ is given in \eqref{eq:ExNT_OR}. Correspondingly, the probability that transmission is failed in previous $n-1$ attempts and $n_j$ is successfully received the packet at the $n^{th}$ attempt\footnote{To make this happen, nodes with a priority higher than $n_j$ must also fail in the $n^{th}$ attempt.} is given by
\begin{align}
\label{eq:P(n)}
\mathcal{P}_{sd}^{ij}(k)=
\begin{cases}
\left(\rm{PER}_{sd}^{i}\right)^{k-1} \overrightarrow{\rm{PDR}}_i^d &, n_j=n_d \\
\left(\rm{PER}_{sd}^{i}\right)^{k-1} \rm{SFR}_{sd}^{ij} &, \text{otherwise}
\end{cases}.
\end{align}
Finally, the EEM fitness of $n_i$ is derived as 
\begin{equation}
\label{eq:EEM_Global}
\rm{F}^i_{\rm{EEM}} \left[\mathcal{C}_{sd}^{i}\right]=\sum_{j \in \mathcal{C}_{sd}^i} \sum_{k=1}^{\rm{K}} \rm{E}_{i}^{j}(k) \mathcal{P}_{sd}^{ij}(k) + \rm{E}_i(K) \left(\rm{PER}_{sd}^{i}\right)^{\rm{K}}
\end{equation}
which is the expected total energy cost of reaching to the destination node through the forwarders in $\mathcal{C}_{sd}^i$
Following from \eqref{eq:EEM_Global}, the candidates can be prioritized by their energy consumption towards the destination node as follows
\begin{equation}
 \label{eq:EEM_Global_priority}
\mathcal{R}(\mathcal{C}_{sd}^{i})=
\{j\vert \rm{EEM_{sd}^{j}} < \rm{EEM_{sd}^{l}}, j<l, \forall (j,l) \in  \mathcal{C}_{sd}^{i}\}.
\end{equation} 
Similar to \eqref{eq:energy_e2e}, we can rewrite global version of \eqref{eq:latency} in a recursive form as 
\begin{equation}
\label{eq:latency_e2e}
\rm{D}_{i}^{j}(n)=\rm{D}_i(n)+\rm{D}_j(\rm{N}_{sd}^j),
\end{equation}
which then can be used to calculate the LLM fitness of $n_i$ as 
\begin{equation}
\label{eq:LLM_Global}
\rm{F}^i_{\rm{LLM}} \left[\mathcal{C}_{sd}^{i}\right]=\sum_{j \in \mathcal{C}_{sd}^i} \sum_{k=1}^{\rm{K}} \rm{D}_{i}^{j}(k) \mathcal{P}_{sd}^{ij}(k) + \rm{D}_i(K) \left(\rm{PER}_{sd}^{i}\right)^{\rm{K}}
\end{equation}
which is the expected total delay to reach the destination node through the forwarders in $\mathcal{C}_{sd}^i$. Following from \eqref{eq:LLM_Global}, the candidates can be prioritized by their energy consumption towards the destination node as follows
\begin{equation}
 \label{eq:LLM_Global_priority}
\mathcal{R}(\mathcal{C}_{sd}^{i})=
\{j\vert \rm{LLM_{sd}^{j}} < \rm{LLM_{sd}^{l}}, j<l, \forall (j,l) \in  \mathcal{C}_{sd}^{i}\}.
\end{equation} 
These E2E metrics can also be transformed into a global ExNT metric as explained at the end of previous section.  

\section{\textbf{SectOR}: \textbf{Sect}or-Based \textbf{O}pportunistic \textbf{R}outing}
\label{sec:SectOR}
In this section, we focus on the design and provision of the proposed SectOR protocol that consists of three main components: 1) Candidate Filtering, 2) Candidate Selection, and 3) Candidate Coordination. Then, we provide algorithmic implementation of the SectOR.

\subsection{Candidate Filtering}
\begin{figure}[!t]
\begin{center}
\includegraphics[width=1\columnwidth]{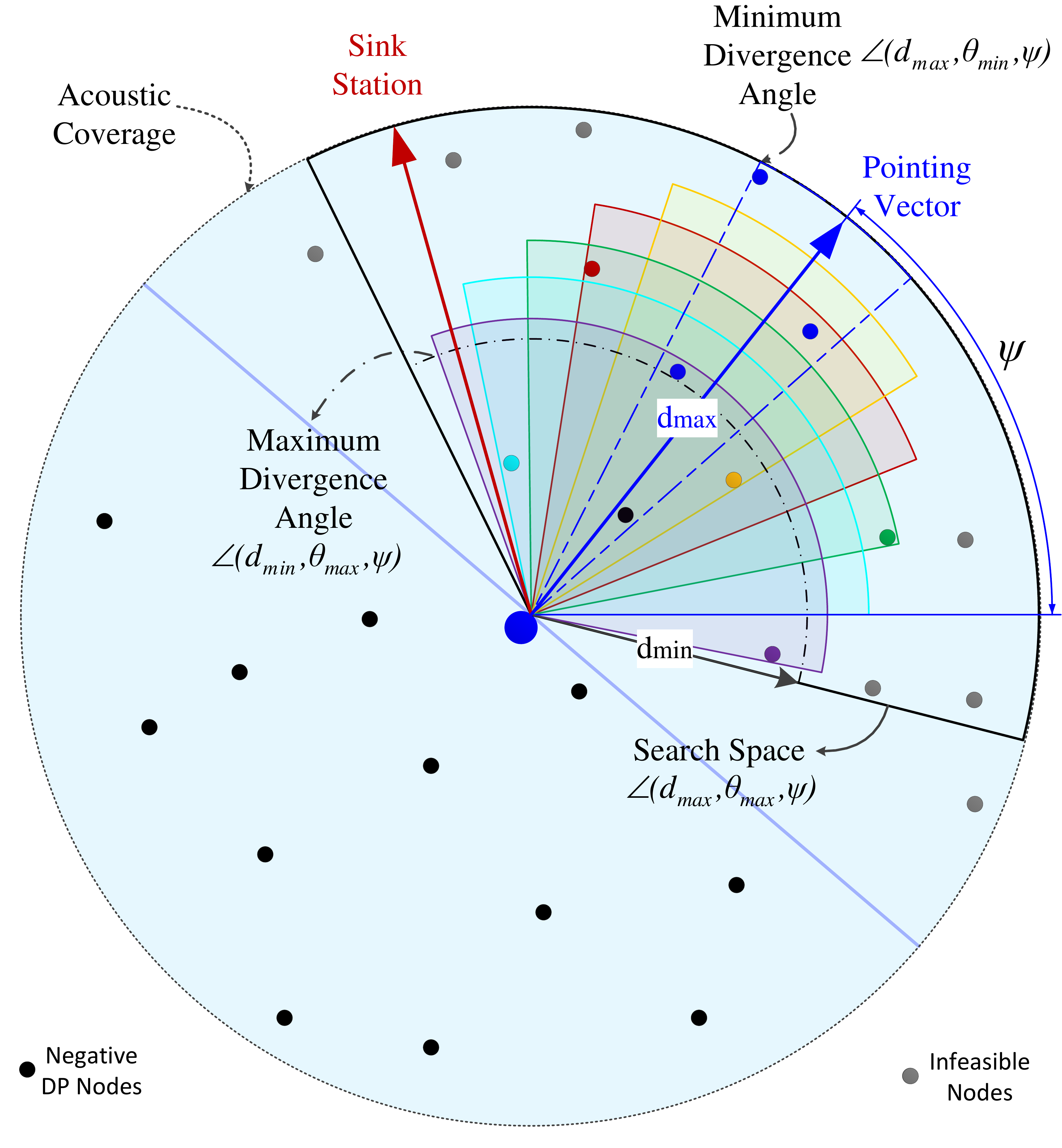}
\caption{Illustration of Candidate Filtering and Selection.}
\vspace{-0.7cm}
\label{fig:CSA}
\end{center}
\end{figure}

The sector-shaped optical coverage region changes with two prominent parameters; pointing direction and divergence angle. Thus, candidate filtering determines the search space (SS) where we manipulate these angles to find CS with the best fitness. To provide a better insight into the candidate filtering, let us pictorially explain it with the help of Fig. \ref{fig:CSA} where the acoustic transmission range is set to the maximum optical communication range. For a given divergence angle range $\theta_{\min}\leq  \theta_{1/e}^i \leq \theta_{\max}$, the maximum and minimum distance  $d_{\max}$ and  $d_{\min}$ can be obtained by substituting  $\theta_{\min}$ and $\theta_{\max}$ into \eqref{eq:dist}, respectively. Since the CS is to be extracted from the SS, we should filter the SS out of the disk-shaped acoustic communication range. Thus, the SS of $n_i$ is given by $$ \mathcal{S}_i=\{\vect{\ell}_x \vert \forall x, \vert\vert\vect{\ell}_x -\vect{\ell}_i\vert\vert>d_{\max}, \vert\vert\vect{\ell}_i -\vect{\ell}_d\vert\vert> \vert\vert\vect{\ell}_x -\vect{\ell}_d\vert\vert  \},$$ which is the set of locations falls within the coverage region with a positive DP [c.f. Fig. \ref{fig:CSA}]. Notice that positive DP condition is crucial especially for local EEM and LLM metrics, where paths may be routed to wrong directions. Since GOR metrics have the global network view, the SS of GOR metrics are allowed to consider all nodes within the acoustic range, i.e., $\mathcal{S}_i=\{\vect{\ell}_x \vert \forall x, \vert\vert\vect{\ell}_x -\vect{\ell}_i\vert\vert>d_{\max} \}$.

In our previous work \cite{Celik2018Sector}, we adopted the conventional method of fixing pointing vector towards the sink station as in \cite{VBF,FBR}. Here, we allow each node to determine its pointing angle within the SS area. To this end, $n_i$ can rotate $\psi_i$ in the counter-clockwise direction and record the angle wherever a new node is entered into the SS. The list of these recorded pointing angles of $n_i$ is denoted by $\Psi_i$. Accordingly, $n_i$ evaluates the fitness of each pointing angle as described in the next section and select the best pointing angle $\psi_i^\star$ as in \eqref{eq:psi}. These two approaches are illustrated in Fig. \ref{fig:CSA} where setting the pointing vector towards the sink station is not desirable because it delivers a poor performance. 

\begin{figure*}
\begin{equation}
\label{eq:psi}
\psi_i^\star= 
\begin{cases}
\underset{\psi \in \Psi_i}{\max} \left( \rm{F}_X^i\left[\mathcal{C}_{sd}^{i \star}(\psi_i) \right] \right)&, \text{ for DP, EDP} \\
\underset{\psi \in \Psi_i}{\min} \left( \rm{F}_X^i\left[\mathcal{C}_{sd}^{i \star}(\psi_i)\right] \right)&, \text{ otherwise}
\end{cases}
\end{equation}

\begin{equation}
\label{eq:CS*}
\mathcal{C}_{sd}^{i\star}(\psi_i)=
\begin{cases}
\left\{\mathcal{C}_{i}^{k}(\psi_i) \left \vert k=\underset{1 \leq \ell \leq \aleph_i}{\argmax} \left\{ \rm{F}_X^i \left[\mathcal{C}_{i}^{\ell}(\psi_i) \right]\right\} \right. \right\}&, \rm{X} \in \text{\{DP, EDP\}} \\
\left\{\mathcal{C}_{i}^{k}(\psi_i)  \left \vert k=\underset{1 \leq \ell \leq \aleph_i}{\argmin} \left\{ \rm{F}_X^i \left[\mathcal{C}_{i}^{\ell}(\psi_i) \right] \right\} \right. \right\}&, \text{otherwise}
\end{cases}
\end{equation}

% \begin{equation}
%  \label{eq:priority}
% \mathcal{PR}_{sd}^{i\star}\left(\mathcal{C}_{sd}^{i\star}(\psi_i^\star)\right)=
% \begin{cases}
% \{k\vert \rm{X_{sd}^{ik}} > \rm{X_{sd}^{il}}, k<l, \forall (k,l) \in  \mathcal{C}_{sd}^{i\star}(\psi_i^\star)\}&, \rm{X} \in \text{\{DP, EDP\}} \\
% \{k\vert \rm{X_{sd}^{ik}} < \rm{X_{sd}^{il}}, k<l, \forall (k,l) \in  \mathcal{C}_{sd}^{i\star}(\psi_i^\star)\}&, \text{otherwise}
% \end{cases}
% \end{equation}
\hrule
\end{figure*}

\subsection{Candidate Selection}

As a result of rate$\leftrightarrow$reliability and range$\leftrightarrow$beam-width tradeoffs, the divergence angle has the main impact on feasible candidate sets (colored sectors in Fig. \ref{fig:CSA}) and their performance in terms of the underlying OR metric. For a data and error rate pair, divergence angle determines the range and thus the CS size, which eventually affects DP, EDP, ExNT, energy consumption, delay, etc.    

Therefore, SectOR manipulates the range$\leftrightarrow$beamwidth tradeoff to obtain a CS which delivers the best OR metric. Now let us consider the pointing angle $\psi_i^\star$ and denote the set of feasible CSs by $\mathcal{CS}(\psi_i^\star)=\{\mathcal{C}_{i}^1(\psi_i^\star),..., \mathcal{C}_{i}^{\aleph_i}(\psi_i^\star) \}$ where $\aleph_i=\vert \Psi_i\vert$. In $\mathcal{CS}(\psi_i^\star)$, the first and the last CSs can be given by $\sphericalangle(d_{\max},\theta_{\min}, \psi_i^\star)$ and $\sphericalangle(d_{\min},\theta_{\max},\psi_i^\star)$ which are shown in black colored dashed and dotted-dashed sectors in Fig. \ref{fig:CSA}, respectively. The rest of CSs may be obtained by quantizing the interval $\theta_{1/e}^i  \in [\theta_{\min},  \theta_{\max}]$ which may be computationally complex for higher resolution. However, this complexity can be augmented by selecting only necessary quantization points based on the node locations within the SS. As shown  by colored sectors in Fig. \ref{fig:CSA}, we widen $\theta_{1/e}^i$ starting from $\theta_{\min}$ and create a new CS whenever a new node is covered by the sector shaped coverage region. Accordingly, the best CS of $n_i$ is determined by \eqref{eq:CS*}. 

% \subsection{Candidate Prioritization}
% The forwarding strategy of $\mathcal{C}_{sd}^{i\star}$ is determined by giving a higher priority to nodes with better OR metrics. Accordingly, candidate prioritization can be formulated as in \eqref{eq:priority} where nodes are sorted in ascending or descending order based on the underlying OR metric. This ordered list can be broadcast to the CS in the packet header such that nodes within the acoustic communication range know CS members and their forwarding priority. 

\subsection{Candidate Coordination}
\label{sec:coordination}
For the candidate coordination, we consider an acknowledgment (ACK) based method where candidates return ACK messages in the priority order embedded in the received packet's header. Slotted Acknowledgment (SA) is one of the first coordination methods \cite{Biswas2004SA} where each candidate sends its ACK with a delay of short interframe space (SIFS) duration, $\tau_{SIFS}$, which yields a coordination delay of $T_c^i=C_{sd}^i\left(\tau_{SIFS}+\tau_{ACK}\right)$ where $\tau_{ACK}$ is the duration of ACK signal. Since SA is vulnerable to the collision of data and ACK packages, a compressed slotted acknowledgment (CSA) method is developed by means of channel sensing techniques \cite{zubow2007multi} that has a coordination delay of $T_c^i=\tau_{SIFS}+C_{sd}^i\tau_{ACK}$. We must note that SectOR is not vulnerable to the collision between data and ACK packages since data and control signals are carried in light and acoustic spectrum, respectively. However, SectOR still needs to account for data transmission of multiple candidates. Hence, we adopt fast slotted acknowledgment (FSA) \cite{Yang2009FSA}, where each candidate waits before deciding whether it should broadcast ACK. This waiting duration is given by $\tau_{SIFS}+(k-1) \tau_{sens}$ where $\tau_{sens}$ is the acoustic channel sensing duration, and $k$ is the priority order in the candidate set. Accordingly, FSA eliminates the possibility of data packet collision at the cost of $T_c^i=\tau_{SIFS}+\tau_{ACK}+C_{sd}^i \: \tau_{sens}$ coordination delay. Noting that $\tau_{sens}$ is negligible in comparison with other terms, FSA requires only a single pair of $\tau_{SIFS}$ and $\tau_{ACK}$.

\begin{algorithm}[t]
 \caption{\small SectOR Protocol}
  \label{alg:sector}
\begin{algorithmic}[1]
 \renewcommand{\algorithmicrequire}{\textbf{Input:}}
 \renewcommand{\algorithmicensure}{\textbf{Output:}}
\Require s, d
\Ensure The best routing path from $n_s$ to $n_d$, i.e., $n_s \rightsquigarrow n_d$
\State $s \rightsquigarrow d$ $\gets$ $n_s$: Initialize the path with $n_s$
\State $n_i \gets n_s$: Set $n_s$ as initial forwarder node
\While{ $n_d \notin s \rightsquigarrow d$ }
\State $\mathcal{R}\left(\mathcal{C}_{sd}^{i\star}(\psi_i^\star)\right) \gets$ \textsc{Filter-Select-Prioritize($\vect{\ell}_i$,$\vect{\ell}_d$)}
\State $n_i$ broadcasts the packet along with $\mathcal{R}\left(\mathcal{C}_{sd}^{i\star}\right)$ 
\State $n_i \gets n_j$: $n_j$ is set as the next forwarder
\State $s \rightsquigarrow d \gets n_j$: Include $n_j$ in the path
\EndWhile
\newline
\Return $s \rightsquigarrow d$
\vspace{3pt}
\hrule 
\vspace{2pt}
\Procedure{Filter-Select-Prioritize}{$\vect{\ell}_i$,$\vect{\ell}_d$}
\State $\mathcal{S}_i \gets$ Determine the search space 
\State $\Psi_i \gets$ Determine the pointing angles 
\For{$\forall \psi_i \in \Psi_i$}
\For{$\forall k \: \vert \: \mathcal{C}_i^k \in \mathcal{CS}(\psi_i)$}
\State $\rm{F}_X^i\left[\mathcal{C}_i^k (\psi_i) \right]  \gets$ Fitness evaluation of metric X 
\EndFor 
\State $\mathcal{C}_{sd}^{i\star}(\psi_i) \gets$ Select the best CS of $\psi_i$ by \eqref{eq:CS*}
\EndFor
\State $\psi_i^\star \gets$ Set the best pointing angle by \eqref{eq:psi}
\State $\mathcal{C}_{sd}^{i\star}(\psi_i^\star) \gets$ Record the best CS by \eqref{eq:CS*}
\State $\mathcal{R}\left(\mathcal{C}_{sd}^{i\star}(\psi_i^\star)\right)\gets$ Prioritize the candidates \\
\hspace{-3pt} \Return $\mathcal{R}\left(\mathcal{C}_{sd}^{i\star}(\psi_i^\star)\right)$
\EndProcedure
 \end{algorithmic}
 \end{algorithm}  

\subsection{Algorithmic Implementation of SectOR}
\label{sec:algorithmic}
Algorithmic implementation of SectOR protocol is summarized in Algorithm \ref{alg:sector} which first initializes the path $n_s \rightsquigarrow n_d$ and current forwarder node to $n_s$. Until $n_d$ is reached, while loop between lines 3 and 8 repeats the following procedure: The current forwarder node $n_i$, execute the candidate filtering, selection, and prioritization procedure by calling \textsc{Filter-Select-Prioritize ($\vect{\ell}_i$)}. In this procedure, lines 10 and 11 calculate the SS, $\mathcal{S}_i$, and set of pointing angles,$\Psi_i$, based on $\vect{\ell}_i$, respectively. Then, the nested for loops between lines 12 and 17 compute OR metrics, evaluates fitness functions, and select the best CS, $\forall \psi_i \in \Psi_i $. Line 18 sets the best pointing angle using \eqref{eq:psi}, then corresponding CSs are calculated using \eqref{eq:CS*} in line 19. Lines 20 ans 21 prioritize the elements of $\mathcal{C}_{sd}^{i\star}$ and return it, respectively. Notice that this procedure is not required to be repeated as long as there is no change in the location vector $\vect{\ell}_i$. Following the procedure, $n_i$ broadcast the packet with the header along with the necessary information such as destination node and priority set. After $T_c^i$ coordination duration, $n_j$ is selected to forward the data packet received from $n_i$. Accordingly, line 6 set $n_j$ as the next forwarder node, and line 7 includes $n_j$ in the path.   

% \begin{figure*}[t]
% \vspace{0.5cm}
% \centering
% \begin{subfigure}[b]{0.24\textwidth}
% \includegraphics[width=\columnwidth]{results/perf_dsp_match_2.eps}
% \caption{}
% \label{fig:a}
% \end{subfigure}
% \begin{subfigure}[b]{0.24\textwidth}
% \includegraphics[width=\columnwidth]{results/rot_search.eps}
%   \caption{}
%   \label{fig:b}
% \end{subfigure}
% \begin{subfigure}[b]{0.24\textwidth}
% \includegraphics[width=\columnwidth]{results/rot2-2.eps}
%   \caption{}
%   \label{fig:c}
%     \end{subfigure}
% \begin{subfigure}[b]{0.24\textwidth}
% \includegraphics[width=\columnwidth]{results/unfeasible.eps}
%   \caption{}
%   \label{fig:d}
%     \end{subfigure}
%     \caption{Illustration of SectOR in comparison with the unicast DSP algorithm.}
%           \label{fig:paths}
% \end{figure*}

% \begin{figure*}[t]
% \centering
% \begin{subfigure}[b]{0.32\textwidth}
% \includegraphics[width=\columnwidth]{results/exnt.eps}
% \caption{Total ExNT}
% \label{fig:exnt}
% \end{subfigure}
% \begin{subfigure}[b]{0.32\textwidth}
% \includegraphics[width=\columnwidth]{results/hops.eps}
%   \caption{Number of Hops}
%   \label{fig:hops}
% \end{subfigure}
% \begin{subfigure}[b]{0.32\textwidth}
% \includegraphics[width=\columnwidth]{results/fails.eps}
%   \caption{Number of Failures}
%   \label{fig:fails}
%     \end{subfigure}
%     \caption{Performance of DP and EDP based SectOR in comparison with the unicast DSP.}
%           \label{fig:SectOR}
% \end{figure*}

\begin{figure*}[t]
\vspace{0.5cm}
\centering
\begin{subfigure}[b]{0.4 \textwidth}
\includegraphics[width=\columnwidth]{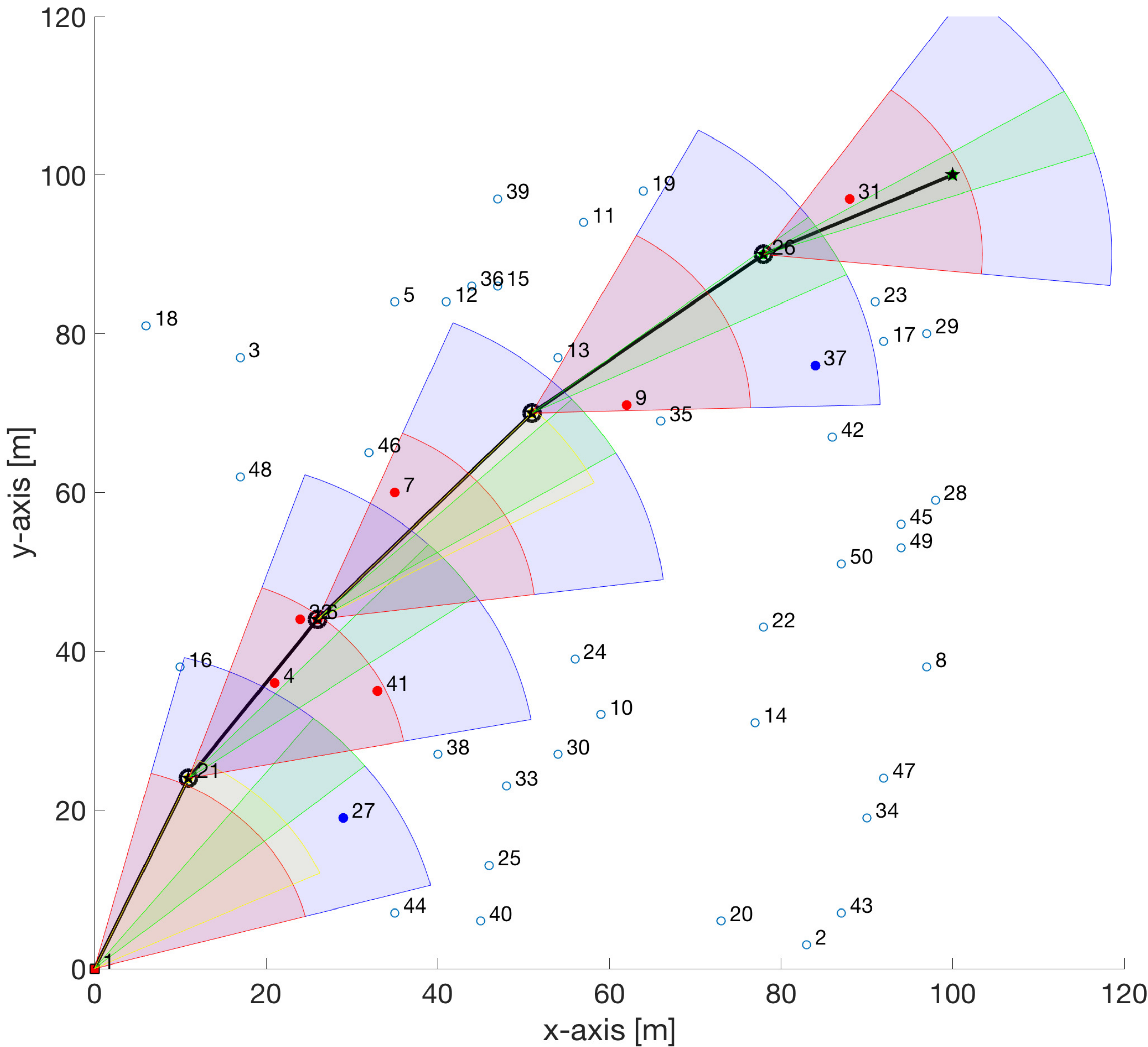}
\caption{}
\label{fig:a}
\end{subfigure}
\begin{subfigure}[b]{0.4 \textwidth}
\includegraphics[width=\columnwidth]{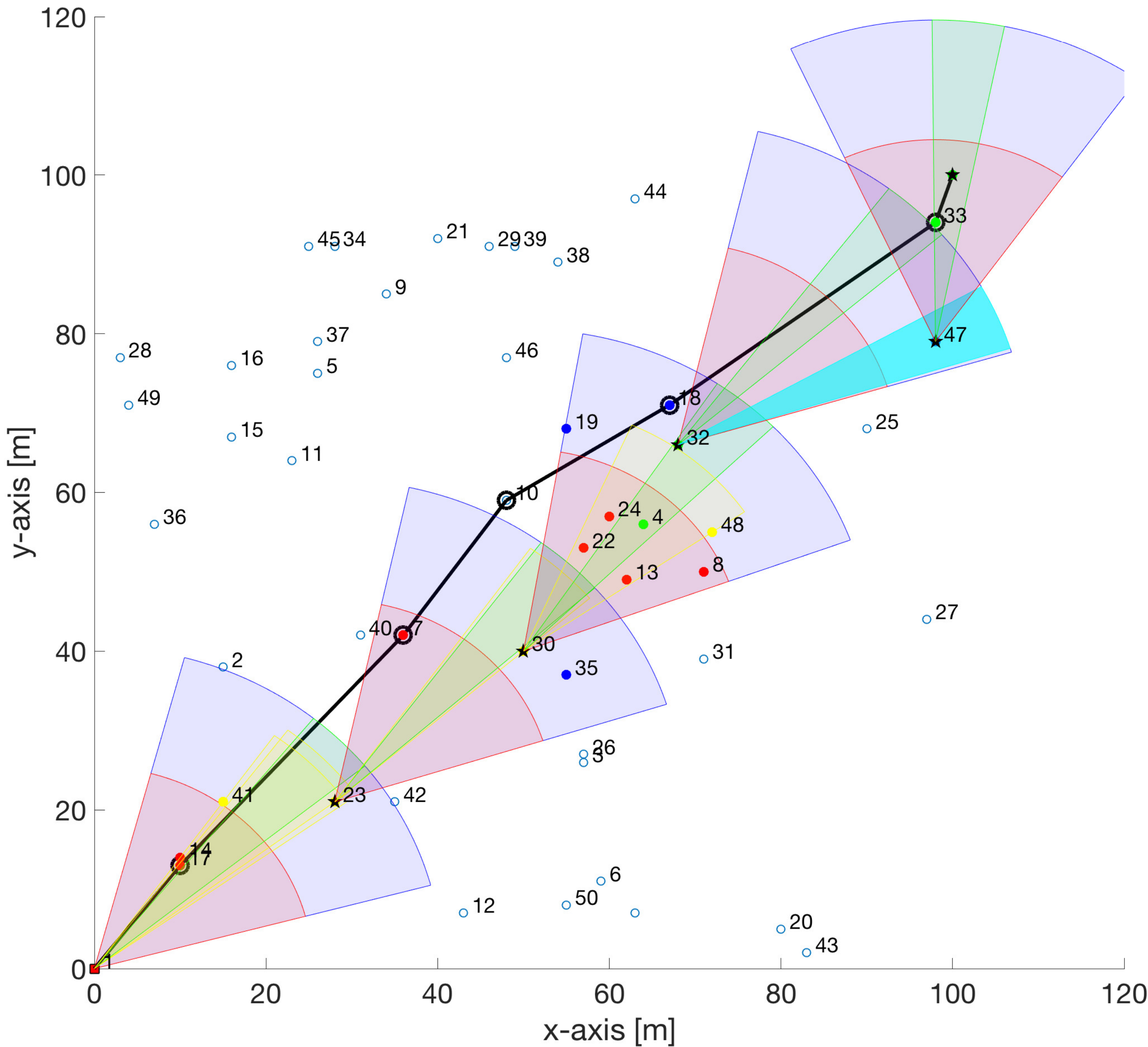}
  \caption{}
  \label{fig:b}
\end{subfigure}

\begin{subfigure}[b]{0.4 \textwidth}
\includegraphics[width=\columnwidth]{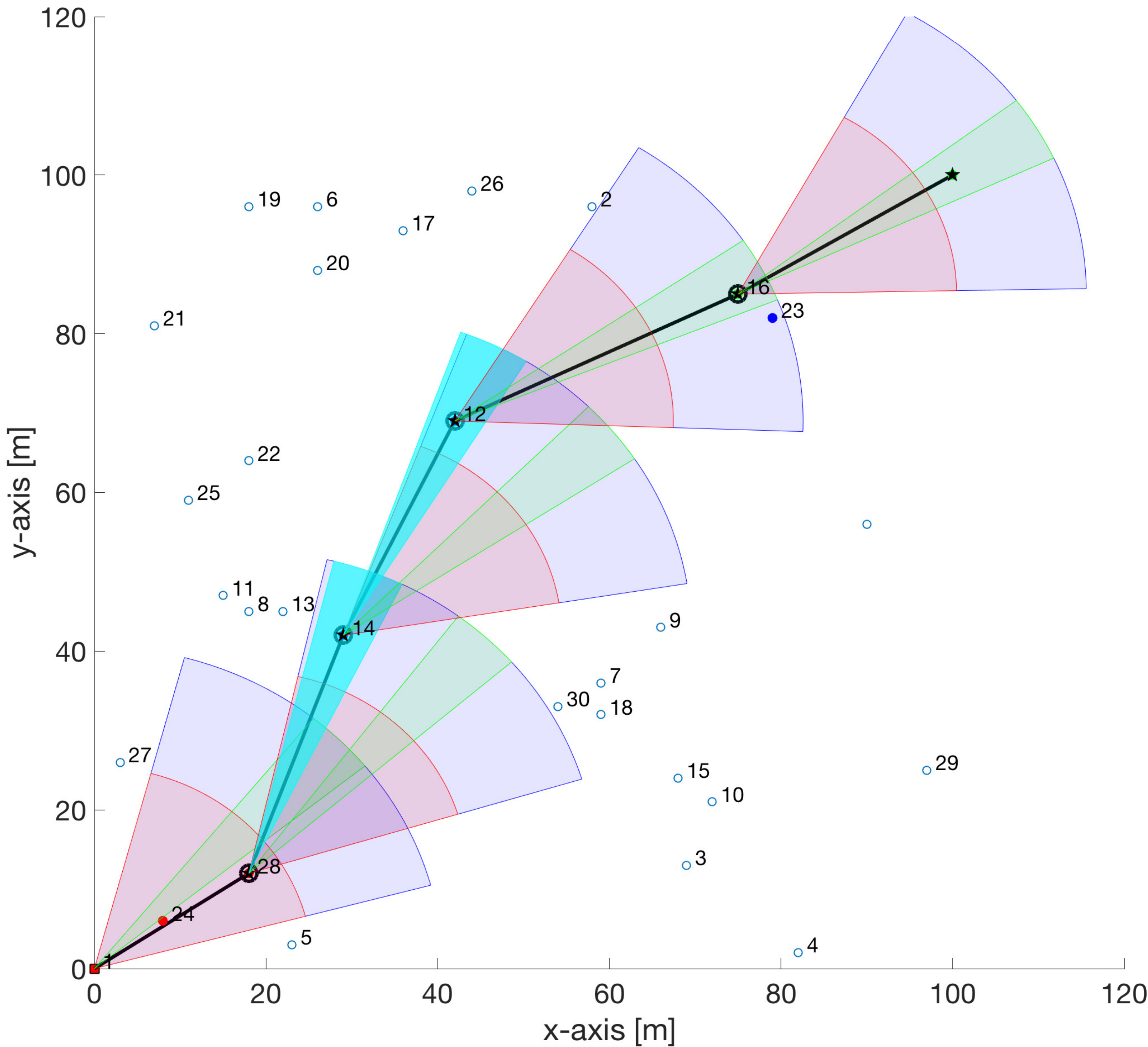}
  \caption{}
  \label{fig:c}
    \end{subfigure}
\begin{subfigure}[b]{0.4 \textwidth}
\includegraphics[width=\columnwidth]{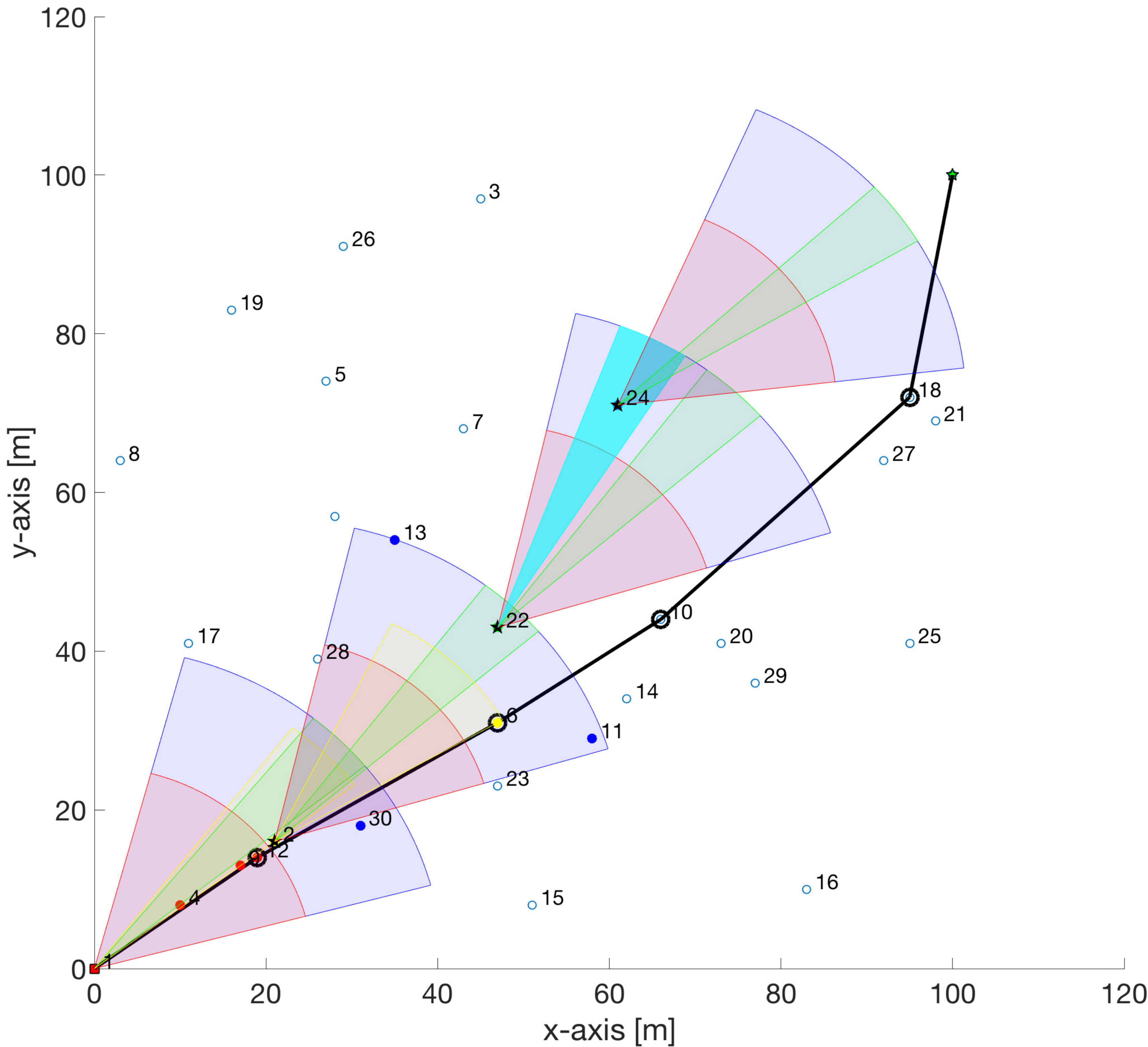}
  \caption{}
  \label{fig:d}
    \end{subfigure}
    \caption{Illustration of SectOR in comparison with TUR.}
          \label{fig:paths}
\end{figure*}

\begin{table}[t]
\centering
\caption{Table of Parameters}
\label{tab:par}
\scriptsize 
\resizebox{\columnwidth}{!}{%
\begin{tabular}{|l|l|l|l|l|l|}
\hline
Par.         & Value       & Par.          & Value           & Par.          & Value         \\ \hline
$Ptx$    & $0.1$ $\rm{W}$ &$\hslash$ & $6.62\mathrm{E}-34$   & $ T$   & $1$ $\rm{ns}$  \\ \hline
$\eta_x$ & $0.9$     &$c$& $2.55\mathrm{E}{8}$ $\rm{m/s}$&$R$      & $1$ $\rm{Gbps}$        \\ \hline
$ \eta_r$& $0.9$     &$\lambda$ & $532\mathrm{E}-9$     & $L$  & $124$ B   \\ \hline
$\eta_c$ & $0.16$    &$e(\lambda)$ & $0.1514$           & $\rm{\overline{PER}}$ & $0.1$     \\ \hline
$A$      & $5$ $\rm{cm^2}$  &$f_{bg}$ & $1\mathrm{E}{6}$       & $f_{dc}$ & $1\mathrm{E}{6}$      \\ \hline
$\delta$      & $0.01$   &$\theta_{\min}$ & $0.336$ rad       & $\theta_{\max}$ & $2/3$ rad      \\ \hline
\end{tabular}%
}
\vspace{-0.6cm}
\end{table}

\section{Numerical Results}
\label{sec:res}
We consider a square network area with varying side lengths (SL). Throughout simulations, the source and the sink nodes are located at reference points of $(0,0)$ and $(\rm{SL,SL})$, respectively. The remaining $50$ nodes are uniformly distributed over the $\rm{SL} \times \rm{SL}$ $m^2$ network area. Obtained results are averaged over $10,000$ random realizations. Unless it is stated explicitly otherwise, we use the parameters listed in Table \ref{tab:par} which is mainly drawn from \cite{Celik2019Endtoend}. Before presenting the numerical results, let us illustrate the operation of SectOR in comparison with a TUR benchmark which is solved  Dijsktra's shortest path (DSP). Unlike the SectOR, the DSP calculates the shortest path based on the available global topology information of the entire network. Therefore, the DSP sets the divergence angle at minimum to reach maximum range with low error performance. 

Before delving into the numerical results, let us provide a better insight into how SectOR protocol works. In Fig. \ref{fig:paths}, black solid lines represent the route calculated by the TUR. Blue/red/green colored sectors correspond to the search space, maximum divergence angle with the minimum range, and minimum divergence angle with the maximum range, respectively. Yellow colored sectors depict the candidate sets obtained by the proposed candidate selection. Fig. \ref{fig:a} shows an instance where the DP metric yields the same path with the TUR benchmark. Fig. \ref{fig:b} is a clear example of how the SectOR can leverage the proposed candidate selection method to find out potential CSs to reach the destination. Notice that the fourth hop starting from node 32 cannot find any feasible CS if the pointing vector is aligned to the sink node. Therefore, node 32 points the transceiver toward the only node (node 47) that lies within the SS, which is shown by cyan colored sector. By doing so, the SectOR was able to reach to the sink node via node 47. A better example of this case is shown in Fig. \ref{fig:c} where the second and third hop is handled by nodes 14 and 12 as there are no nodes within the feasible region of coverage. Finally, Fig. \ref{fig:d} demonstrate the negative impact of the lack of global topology information on the routing performance. Even though the path is routed over node 24 by changing the pointing vector node 22, the local SectOR was not able to reach to the destination.

\begin{figure}[t]
\centering
\includegraphics[width=0.5 \textwidth]{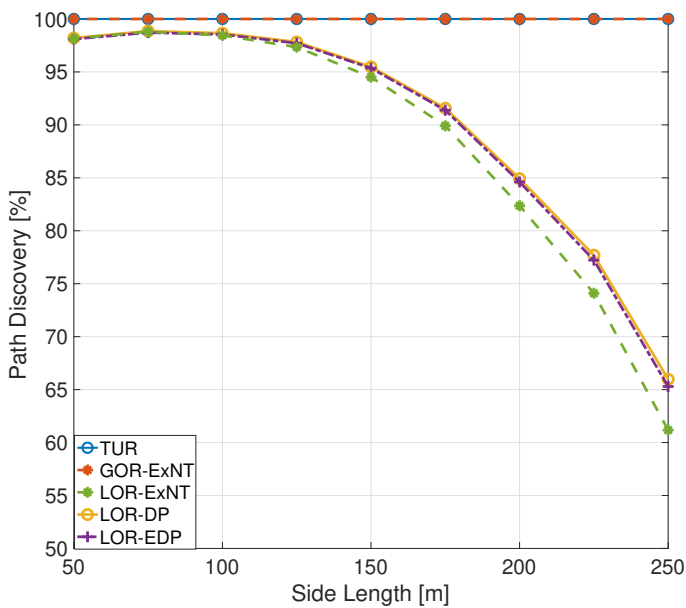}
\caption{Percentage of path discovery vs. side length (density).}
\label{fig:discovery}
\end{figure}

In the remainder, we present key performance metrics for various network side length, i.e., network size. Since we keep number of nodes constant, increasing side lengths also account for decreasing the node density. Fig. \ref{fig:discovery} demonstrate the impacts of node density on the probability of finding a path from source to destination. Since TUR and GOR-EXNT have a global network view, they were always able to find a feasible path. On the contrary, LOR methods have a lower probability of path discovery as the node density decreases. In particular, LOR-DP and LOR-EDP perform slightly better than LOR-ExNT because they try to reach the destination in a lower number of hops, which naturally helps them find a feasible path. On the contrary, LOR-ExNT fails to reach the destination while following paths with less number of ExNT.

\begin{figure}[t]
\centering
\includegraphics[width=0.5 \textwidth]{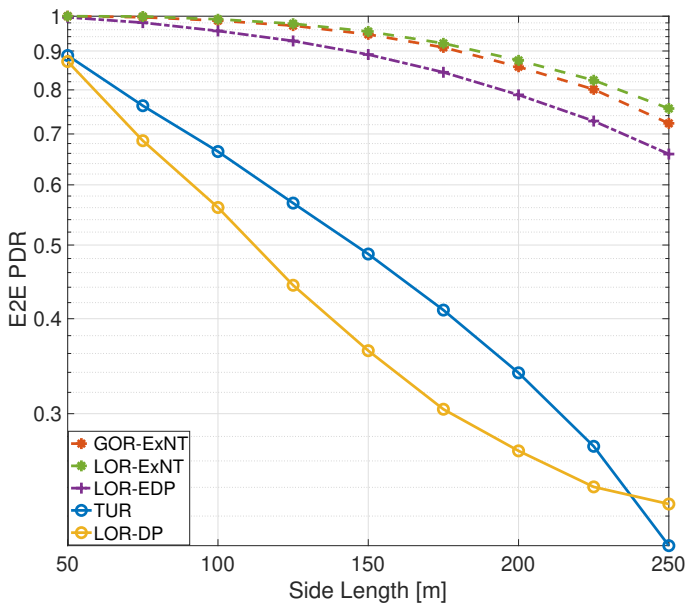}
\caption{E2E-PDR vs. side length (density).}
\label{fig:pdr}
\end{figure}

\begin{figure}[t]
\centering
\includegraphics[width=0.5 \textwidth]{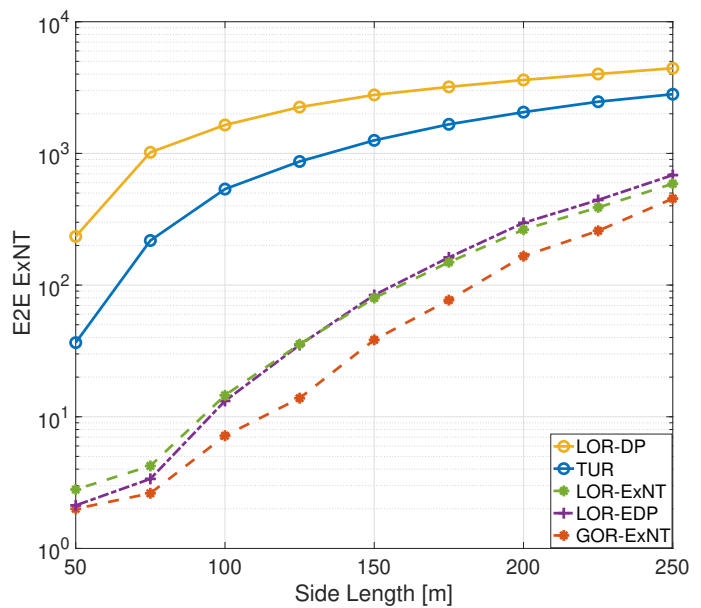}
\caption{E2E-ExNT vs. side length (density).}
\label{fig:exnt}
\end{figure}

Fig. \ref{fig:pdr} demonstrates the fact that E2E-PDR reduces with increasing network sparsity. The first thing to observe is that TUR and LOR-DP perform worse than others, which is mainly because of the fact that they are agnostic to the underwater channel hostility and merely consider the distance as their performance metric. By considering the channel condition along with the distance, LOR-EDP delivers a much better performance than its counterpart LOR-DP. On the other hand, LOR-ExNT delivers a higher PDR as ExNT metric is inversely proportional to the PDR. Lastly, GOR-ExNT is the best as it has the global network view. Fig. \ref{fig:exnt} visualizes the fact that E2E-ExNT increases with decreasing network density. Again, TUR and LOR-DP delivers a poor performance as they only account for the distance. On the contrary, LOR-EDP and LOR-EDP improves the E2E-ExNT, which is beaten by the GOR-ExNT at the cost of a higher computational complexity and communication overhead.

\begin{figure}[t]
\centering
\includegraphics[width=0.5 \textwidth]{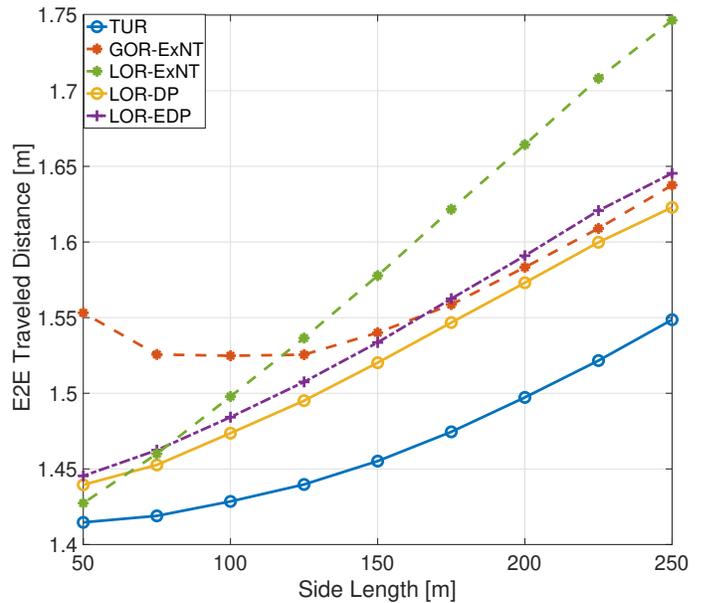}
\caption{E2E-distance vs. side length (density).}
\label{fig:distance}
\end{figure}

Fig. \ref{fig:distance} compares the traveled total distance by different schemes. As expected, TUR gives the shortest distance as the DSP is optimal. TUR is followed by LOR-DP and then LOR-EDP as they account for distance progress. However, LOR-ExNT performs worst as they follow a longer path with higher number of hops for the sake of a minimum ExNT. Although GOR-EXNT is the worst case under higher densities, it reaches a performance level closer to LOR-DP and LOR-EDP.

\begin{figure}[t]
\centering
\includegraphics[width=0.5 \textwidth]{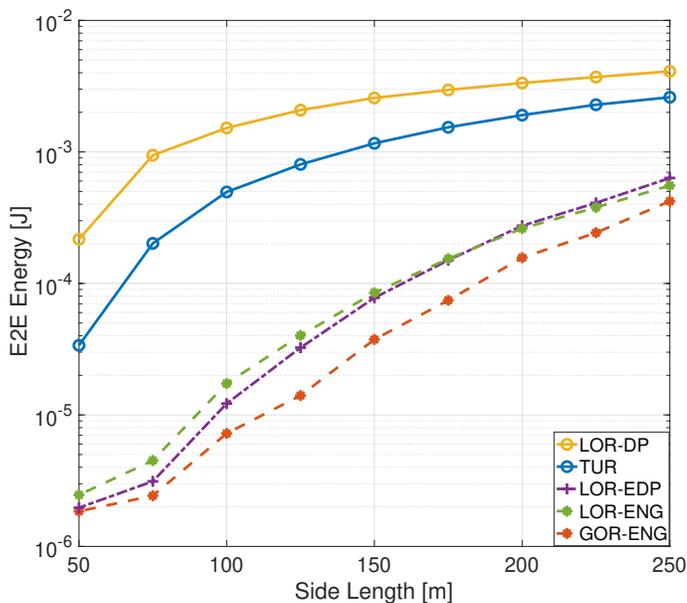}
\caption{E2E-Energy vs. side length (density).}
\label{fig:energy}
\end{figure}

\begin{figure}[t]
\centering
\includegraphics[width=0.5 \textwidth]{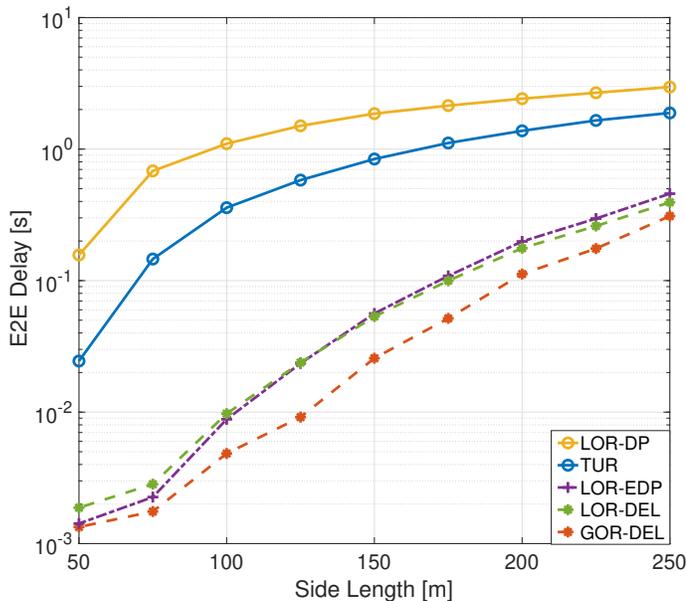}
\caption{E2E-Delay vs. side length (density).}
\label{fig:delay}
\end{figure}

Fig. \ref{fig:energy} and Fig. \ref{fig:delay} show the E2E energy consumption and delay versus varying network size. The LOR-DP yields highest energy consumption and delay for the sake of the shortest path. The TUR provides second worst case performance. LOR-EDP and LOR-ExNT improves the energy and delay performance significantly, which is again beaten by the GOR-ENG and GOR-DEL, respectively.

% \begin{figure*}[t]
% \centering
% \begin{subfigure}[b]{0.32\textwidth}
% \includegraphics[width=\columnwidth]{results/discovery.eps}
% \label{fig:discovery}
% \caption{Total ExNT}
% \end{subfigure}
% \begin{subfigure}[b]{0.32\textwidth}
% \includegraphics[width=\columnwidth]{results/pdr.eps}
% \label{fig:pdr}
% \caption{}
% \end{subfigure} 
% \begin{subfigure}[b]{0.32\textwidth}
% \includegraphics[width=\columnwidth]{results/distance.eps}
% \label{fig:distance}
% \caption{}
% \end{subfigure} 
% \caption{Performance of DP and EDP based SectOR in comparison with the unicast DSP.}
% \label{fig:SectOR}
% \end{figure*}

% \begin{figure*}[t]
% \centering
% \begin{subfigure}[b]{0.32\textwidth}
% \includegraphics[width=\columnwidth]{results/exnt.eps}
% \label{fig:exnt}
% \caption{}
% \end{subfigure}
% \begin{subfigure}[b]{0.32\textwidth}
% \includegraphics[width=\columnwidth]{results/energy.eps}
% \label{fig:energy}
% \caption{}
% \end{subfigure}
% \begin{subfigure}[b]{0.32\textwidth}
% \includegraphics[width=\columnwidth]{results/delay.eps}
% \label{fig:delay}
% \caption{}
% \end{subfigure}
% \caption{Performance of DP and EDP based SectOR in comparison with the unicast DSP.}
% \label{fig:SectOR}
% \end{figure*}

\section{Conclusions}
\label{sec:conc}
In this paper, we developed a sector-based opportunistic routing protocol for UOANs. OR is especially suitable for UOWCs as hostile aquatic channel impairments can disrupt the established link connectivity. By leveraging the broadcast nature of the communication, backing up the broken link connectivity by engaging other users who also received the lost packet may improve the system performance in a significant extend. The proposed SectOR protocol exploits the sector-shaped coverage region of the light source and finds the path exploiting the local or global topology information in a distributed manner. Numerical results show that it can reach the performance level of a unicast optimal routing especially for high node density levels.

%\appendix
%\section{Table of Symbols}

% References
\bibliographystyle{IEEEtran}

\bibliography{iotj}

% Generated by IEEEtran.bst, version: 1.14 (2015/08/26)
\begin{thebibliography}{10}
\providecommand{\url}[1]{#1}
\csname url@samestyle\endcsname
\providecommand{\newblock}{\relax}
\providecommand{\bibinfo}[2]{#2}
\providecommand{\BIBentrySTDinterwordspacing}{\spaceskip=0pt\relax}
\providecommand{\BIBentryALTinterwordstretchfactor}{4}
\providecommand{\BIBentryALTinterwordspacing}{\spaceskip=\fontdimen2\font plus
\BIBentryALTinterwordstretchfactor\fontdimen3\font minus
  \fontdimen4\font\relax}
\providecommand{\BIBforeignlanguage}[2]{{%
\expandafter\ifx\csname l@#1\endcsname\relax
\typeout{** WARNING: IEEEtran.bst: No hyphenation pattern has been}%
\typeout{** loaded for the language `#1'. Using the pattern for}%
\typeout{** the default language instead.}%
\else
\language=\csname l@#1\endcsname
\fi
#2}}
\providecommand{\BIBdecl}{\relax}
\BIBdecl

\bibitem{Celik2018Sector}
A.~Celik, N.~Saeed, B.~Shihada, T.~Y. Al-Naffouri, and M.-S. Alouini,
  ``Sect{OR}: Sector-based opportunistic routing protocol for underwater
  optical wireless networks,'' in \emph{IEEE Wireless Commun. Netw. Conf.
  (WCNC)}, Apr. 2019, pp. 1--6.

\bibitem{Celik2018Survey}
N.~{Saeed}, A.~{Celik}, T.~Y. {Al-Naffouri}, and M.-S. {Alouini}, ``Underwater
  optical wireless communications, networking, and localization: {A} survey,''
  \emph{Ad Hoc Networks}, vol.~94, p. 101935, 2019.

\bibitem{celik2019software}
A.~Celik, N.~Saeed, B.~Shihada, T.~Y. Al-Naffouri, and M.-S. Alouini, ``A
  software-defined opto-acoustic network architecture for internet of
  underwater things,'' \emph{arXiv preprint arXiv:1910.05306}, 2019.

\bibitem{Melike2011Survey}
M.~{Erol-Kantarci}, H.~T. {Mouftah}, and S.~{Oktug}, ``A survey of
  architectures and localization techniques for underwater acoustic sensor
  networks,'' \emph{IEEE Communications Surveys Tutorials}, vol.~13, no.~3, pp.
  487--502, Third 2011.

\bibitem{Celik2018Modeling}
A.~Celik, N.~Saeed, T.~Y. Al-Naffouri, and M.-S. Alouini, ``Modeling and
  performance analysis of multihop underwater optical wireless sensor
  networks,'' in \emph{IEEE Wireless Commun. Netw. Conf. (WCNC)}, Apr. 2018,
  pp. 1--6.

\bibitem{Celik2019Endtoend}
A.~{Celik}, N.~{Saeed}, T.~Y. {Al-Naffouri}, and M.-S. {Alouini}, ``End-to-end
  performance analysis of underwater optical wireless relaying and routing
  techniques under location uncertainty,'' \emph{IEEE Trans. Wireless Commun.},
  2019.

\bibitem{DOMINGO2012IouT}
M.~C. Domingo, ``An overview of the internet of underwater things,''
  \emph{Journal of Network and Computer Applications}, vol.~35, no.~6, pp. 1879
  -- 1890, 2012.

\bibitem{Kao2017IoUT}
C.-C. Kao, Y.-S. Lin, G.-D. Wu, and C.-J. Huang, ``A comprehensive study on the
  internet of underwater things: Applications, challenges, and channel
  models,'' \emph{Sensors}, vol.~17, no.~7, 2017.

\bibitem{AKYILDIZ20161}
I.~F. Akyildiz, P.~Wang, and S.-C. Lin, ``Softwater: Software-defined
  networking for next-generation underwater communication systems,'' \emph{Ad
  Hoc Networks}, vol.~46, pp. 1 -- 11, 2016.

\bibitem{Celik2018Connectivity}
N.~Saeed, A.~Celik, T.~Y. Al-Naffouri, and M.-S. Alouini, ``Connectivity
  analysis of underwater optical wireless sensor networks: A graph theoretic
  approach,'' in \emph{IEEE Int. Conf. Commun. Workshops (ICC Workshops)}, May
  2018, pp. 1--6.

\bibitem{Jamali2016performance}
M.~V. Jamali, F.~Akhoundi, and J.~A. Salehi, ``Performance characterization of
  relay-assisted wireless optical {CDMA} networks in turbulent underwater
  channel,'' \emph{IEEE Trans. Wireless Commun.}, vol.~15, no.~6, pp.
  4104--4116, June 2016.

\bibitem{Jamali2017multihop}
M.~V. Jamali, A.~Chizari, and J.~A. Salehi, ``Performance analysis of multi-hop
  underwater wireless optical communication systems,'' \emph{IEEE Photon.
  Technol. Lett.}, vol.~29, no.~5, pp. 462--465, March 2017.

\bibitem{Chakchouk2015SurveyOR}
N.~{Chakchouk}, ``A survey on opportunistic routing in wireless communication
  networks,'' \emph{IEEE Communications Surveys Tutorials}, vol.~17, no.~4, pp.
  2214--2241, Fourthquarter 2015.

\bibitem{Menon2016Comparative}
V.~G. {Menon} and P.~M.~J. {Prathap}, ``Comparative analysis of opportunistic
  routing protocols for underwater acoustic sensor networks,'' in \emph{Intl.
  Conf. Emerging Techno. Trends (ICETT)}, Oct 2016, pp. 1--5.

\bibitem{Darehshoorzadeh2015Underwater}
A.~{Darehshoorzadeh} and A.~{Boukerche}, ``Underwater sensor networks: a new
  challenge for opportunistic routing protocols,'' \emph{IEEE Commun. Mag.},
  vol.~53, no.~11, pp. 98--107, Nov. 2015.

\bibitem{Coutinho2016Geographic}
R.~W.~L. {Coutinho}, A.~{Boukerche}, L.~F.~M. {Vieira}, and A.~A.~F.
  {Loureiro}, ``Geographic and opportunistic routing for underwater sensor
  networks,'' \emph{IEEE Trans. Computers}, vol.~65, no.~2, pp. 548--561, Feb.
  2016.

\bibitem{Coutinho2014GEDAR}
------, ``{GEDAR}: Geographic and opportunistic routing protocol with depth
  adjustment for mobile underwater sensor networks,'' in \emph{IEEE Intl. Conf.
  Commun. (ICC)}, Jun. 2014, pp. 251--256.

\bibitem{LoPresti06}
P.~LoPresti, H.~Refai, J.~Sluss, and I.~Varela-Cuadrado, ``Adaptive divergence
  and power for improving connectivity in free-space optical mobile networks,''
  \emph{Appl. Opt.}, vol.~45, no.~25, pp. 6591--6597, Sep. 2006.

\bibitem{Celik2019Localization}
N.~{Saeed}, A.~{Celik}, T.~Y. {Al-Naffouri}, and M.-S. {Alouini},
  ``Localization of energy harvesting empowered underwater optical wireless
  sensor networks,'' \emph{IEEE Trans. Wireless Commun.}, vol.~18, no.~5, pp.
  2652--2663, May 2019.

\bibitem{Celik2019Performance}
------, ``Performance analysis of connectivity and localization in multi-hop
  underwater optical wireless sensor networks,'' \emph{IEEE Trans. Mobile
  Comput.}, vol.~18, no.~11, pp. 2604--2615, Nov 2019.

\bibitem{Celik2017HybAcoOpt}
N.~Saeed, A.~Celik, M.-S. Alouini, and T.~Y. Al-Naffouri, ``Energy harvesting
  hybrid acoustic-optical underwater wireless sensor networks localization,''
  \emph{Sensors}, vol.~18, no.~1, 2017.

\bibitem{Poliak2012link}
J.~Poliak, P.~Pezzei, E.~Leitgeb, and O.~Wilfert, ``Link budget for high-speed
  short-distance wireless optical link,'' in \emph{2012 8th International
  Symposium on Communication Systems, Networks Digital Signal Processing
  (CSNDSP)}, Jul. 2012, pp. 1--6.

\bibitem{Elamassie2018performance}
M.~Elamassie, F.~Miramirkhani, and M.~Uysal, ``Performance characterization of
  underwater visible light communication,'' \emph{IEEE Trans. Commun.}, pp.
  1--1, 2018.

\bibitem{Arnon10underwater}
S.~Arnon, ``Underwater optical wireless communication network,'' \emph{Optical
  Engineering}, vol.~49, pp. 49 -- 49 -- 6, 2010.

\bibitem{ZorziDP2003}
M.~Zorzi and R.~R. Rao, ``Geographic random forwarding ({G}e{R}a{F}) for ad hoc
  and sensor networks: multihop performance,'' \emph{IEEE Trans. Mobile
  Comput.}, vol.~2, no.~4, pp. 337--348, Oct. 2003.

\bibitem{Darehshoorzadeh2012distance}
A.~Darehshoorzadeh and L.~Cerdà-Alabern, ``Distance progress based
  opportunistic routing for wireless mesh networks,'' in \emph{2012 8th
  International Wireless Communications and Mobile Computing Conference
  (IWCMC)}, Aug 2012, pp. 179--184.

\bibitem{VBF}
P.~Xie, J.-H. Cui, and L.~Lao, ``{VBF}: Vector-based forwarding protocol for
  underwater sensor networks,'' in \emph{NETWORKING 2006. Networking
  Technologies, Services, and Protocols; Performance of Computer and
  Communication Networks; Mobile and Wireless Communications Systems},
  F.~Boavida, T.~Plagemann, B.~Stiller, C.~Westphal, and E.~Monteiro,
  Eds.\hskip 1em plus 0.5em minus 0.4em\relax Berlin, Heidelberg: Springer
  Berlin Heidelberg, 2006, pp. 1216--1221.

\bibitem{FBR}
J.~M. Jornet, M.~Stojanovic, and M.~Zorzi, ``Focused beam routing protocol for
  underwater acoustic networks,'' in \emph{Proceedings of the Third ACM
  International Workshop on Underwater Networks}, ser. WuWNeT '08.\hskip 1em
  plus 0.5em minus 0.4em\relax New York, NY, USA: ACM, 2008, pp. 75--82.

\bibitem{Biswas2004SA}
S.~Biswas and R.~Morris, ``Opportunistic routing in multi-hop wireless
  networks,'' \emph{SIGCOMM Comput. Commun. Rev.}, vol.~34, no.~1, pp. 69--74,
  2004.

\bibitem{zubow2007multi}
A.~Zubow, M.~Kurth, and J.-P. Redlich, ``Multi-channel opportunistic routing,''
  in \emph{European Wireless Conference (EW)}, 2007.

\bibitem{Yang2009FSA}
Z.~{Yang}, K.~{Zeng}, and W.~{Lou}, ``{FSA}: A fast coordination scheme for
  opportunistic routing,'' in \emph{IEEE Intl. Conf. Commun. (ICC)}, Jun 2009,
  pp. 1--5.

\end{thebibliography}
\vspace*{-2\baselineskip}
 \begin{IEEEbiography}[{\includegraphics[width=1.1 in,height=1.25in]{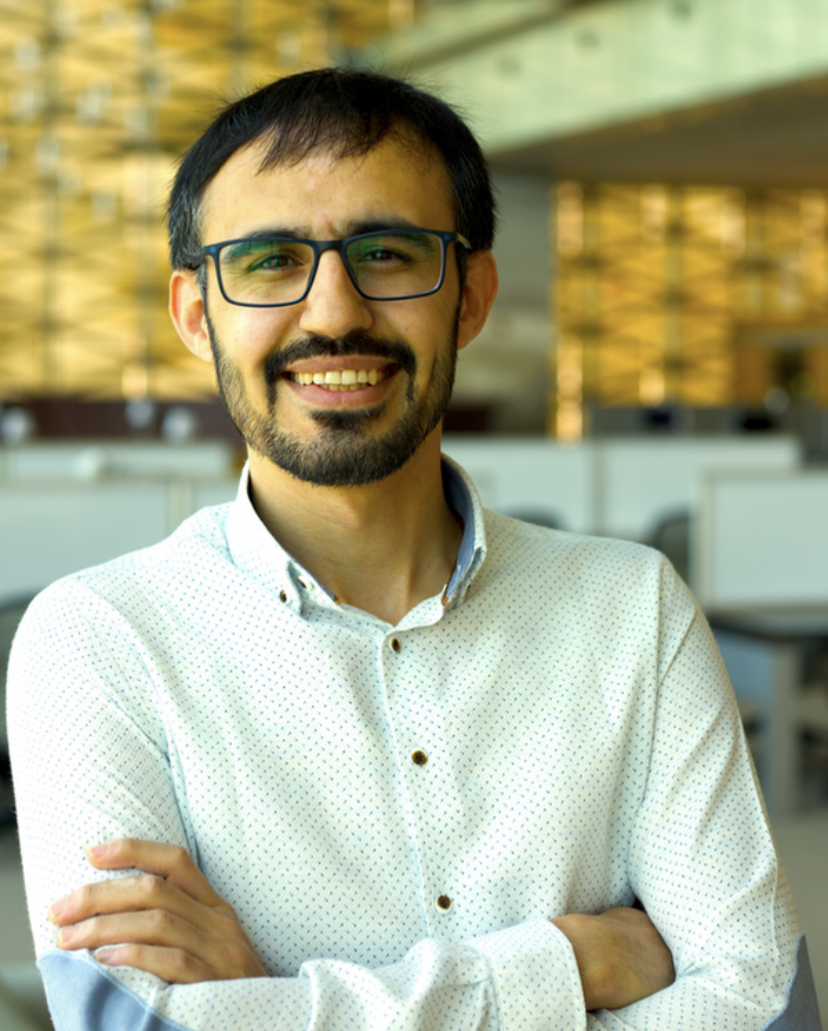}}]{Abdulkadir Celik}
 (S'14-M'16-SM'19) received the B.S. degree in electrical-electronics engineering from Selcuk University in 2009, the M.S. degree in electrical engineering in 2013, the M.S. degree in computer engineering in 2015, and the Ph.D. degree in co-majors of electrical engineering and computer engineering in 2016, all from Iowa State University, Ames, IA. He is currently a postdoctoral research fellow at Communication Theory Laboratory of King Abdullah University of Science and Technology (KAUST). His current research interests include but not limited to 5G networks and beyond, wireless data centers, UAV assisted cellular and IoT networks, and underwater optical wireless communications, networking, and localization. 
 \end{IEEEbiography}

 \vspace*{-2\baselineskip}

\begin{IEEEbiography}[{\includegraphics[width=1.1 in,height=1.25in]{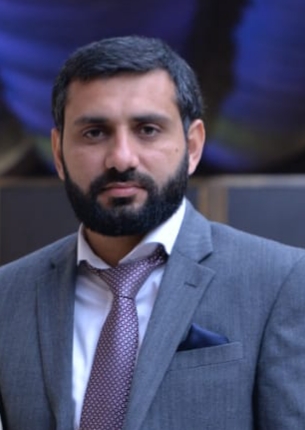}}]{Nasir Saeed}(S'14-M'16-SM'19) received his Bachelors of Telecommunication degree from University of Engineering and Technology, Peshawar, Pakistan, in 2009 and received Masters degree in satellite navigation from Polito di Torino, Italy, in 2012. He received his Ph.D. degree in electronics and communication engineering from Hanyang University, Seoul, South Korea in 2015. He was an assistant professor at the Department of Electrical Engineering, Gandhara Institute of Science and IT, Peshawar, Pakistan from August 2015 to September 2016. Dr. Saeed worked as an assistant professor at IQRA National University, Peshawar, Pakistan from October 2017 to July 2017. He is currently a postdoctoral research fellow at Communication Theory Lab, King Abdullah University of Science and Technology (KAUST).   His current areas of interest include cognitive radio networks, underwater optical wireless communications, dimensionality reduction, and localization.
 \end{IEEEbiography}

 \vspace*{-2\baselineskip}
 \begin{IEEEbiography}[{\includegraphics[width=1.1in,height=1.25in]{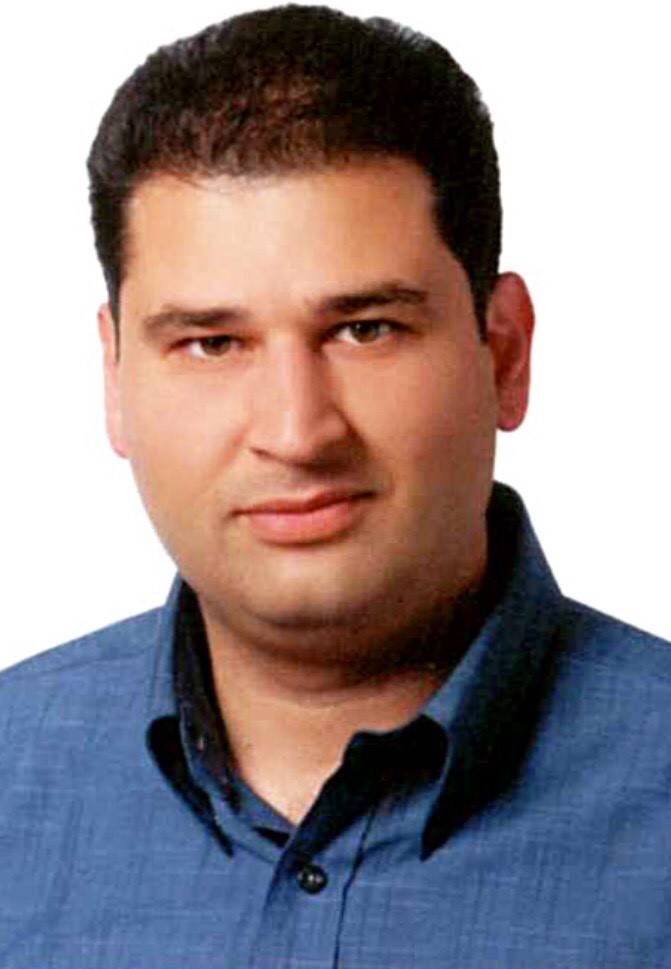}}]{Basem Shihada}(SM'12)  is an associate \& founding professor in the Computer, Electrical and Mathematical Sciences \& Engineering (CEMSE) Division at King Abdullah University of Science and Technology (KAUST). He obtained his PhD in Computer Science from University of Waterloo. In 2009, he was appointed as visiting faculty in the Department of Computer Science, Stanford University. In 2012, he was elevated to the rank of Senior Member of IEEE. His current research covers a range of topics in energy and resource allocation in wired and wireless networks, software defined networking, internet of things, data networks, network security, and cloud/fog computing. 
 \end{IEEEbiography}
 \vspace*{-2\baselineskip}

\begin{IEEEbiography}[{\includegraphics[width=1.1in,height=1.25in]{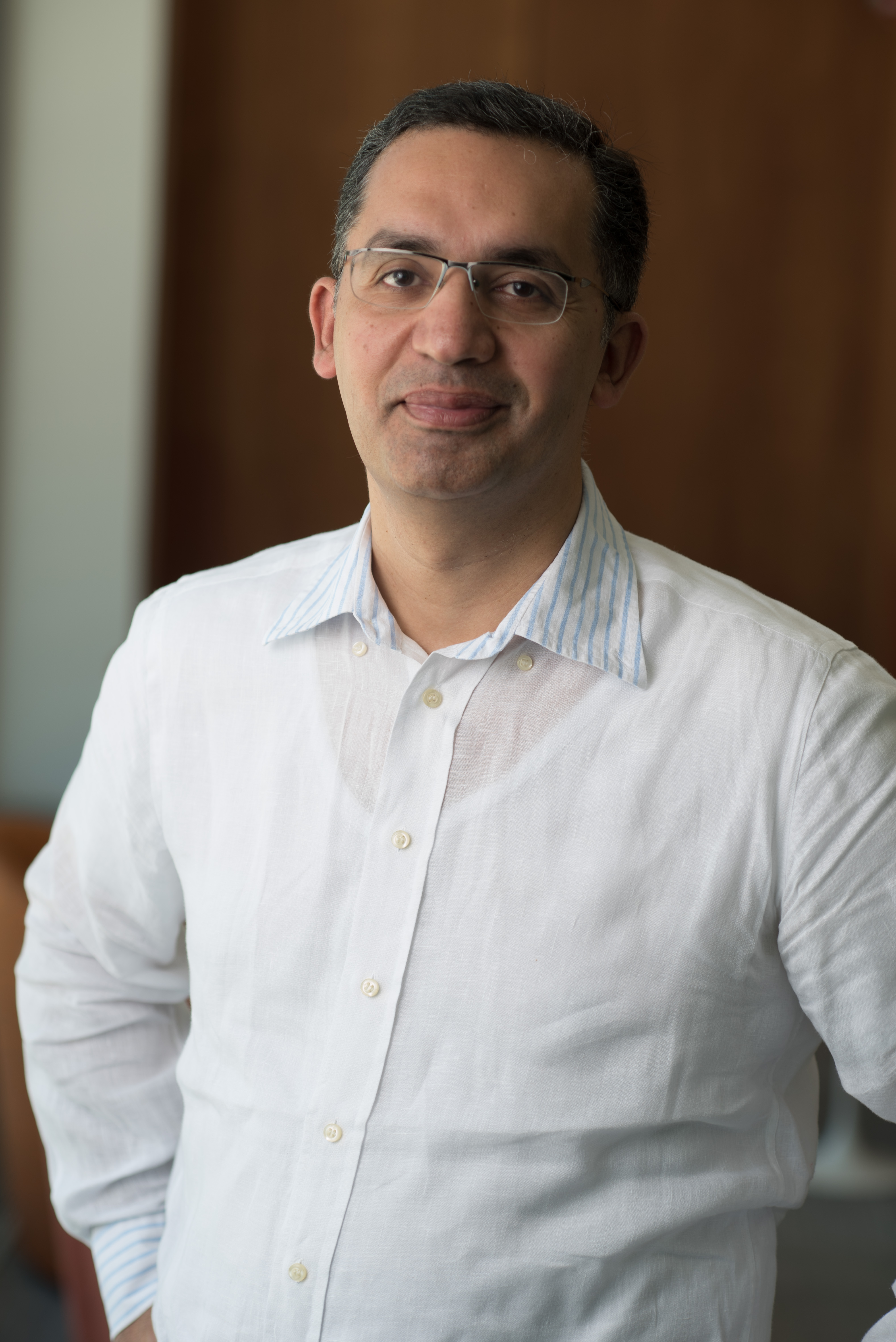}}]{Tareq Y. Al-Naffouri }
 (M'10-SM'18) Tareq  Al-Naffouri  received  the  B.S.  degrees  in  mathematics  and  electrical  engineering  (with  first  honors)  from  King  Fahd  University  of  Petroleum  and  Minerals,  Dhahran,  Saudi  Arabia,  the  M.S.  degree  in  electrical  engineering  from  the  Georgia  Institute  of  Technology,  Atlanta,  in  1998,  and  the  Ph.D.  degree  in  electrical  engineering  from  Stanford  University,  Stanford,  CA,  in  2004.  

 He  was  a  visiting  scholar  at  California  Institute  of  Technology,  Pasadena,  CA  in  2005  and  summer  2006.  He  was  a  Fulbright scholar  at  the  University  of  Southern  California  in  2008.  He  has  held  internship  positions  at  NEC  Research  Labs,  Tokyo,  Japan,  in  1998,  Adaptive  Systems  Lab,  University  of  California  at  Los  Angeles  in  1999,  National  Semiconductor,  Santa  Clara,  CA,  in  2001  and  2002,  and  Beceem  Communications  Santa  Clara,  CA,  in  2004.  He  is  currently  an  Associate Professor  at  the  Electrical  Engineering  Department,  King  Abdullah  University  of  Science  and  Technology  (KAUST).  His  research  interests  lie  in  the  areas  of  sparse, adaptive,  and  statistical  signal  processing  and  their  applications,  localization,  machine  learning,  and  network  information  theory.    He  has  over  240  publications  in  journal  and  conference  proceedings,  9  standard  contributions,  14  issued  patents,  and  8  pending. 

 Dr.  Al-Naffouri  is  the  recipient  of  the  IEEE  Education  Society  Chapter  Achievement  Award  in  2008  and  Al-Marai  Award  for  innovative  research  in  communication  in  2009.  Dr.  Al-Naffouri  has  also  been  serving  as  an  Associate  Editor  of  Transactions  on  Signal  Processing  since  August  2013. 
 \end{IEEEbiography}
 \vspace*{-2\baselineskip}

 \begin{IEEEbiography}[{\includegraphics[width=1.1in,height=1.25in]{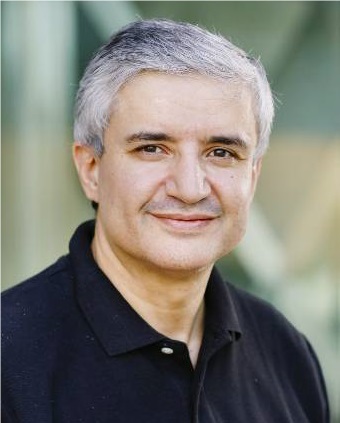}}]{Mohamed-Slim Alouini}
 (S'94-M'98-SM'03-F'09)  was born in Tunis, Tunisia. He received the Ph.D. degree in Electrical Engineering
 from the California Institute of Technology (Caltech), Pasadena,
 CA, USA, in 1998. He served as a faculty member in the University of Minnesota,
 Minneapolis, MN, USA, then in the Texas A\&M University at Qatar,
 Education City, Doha, Qatar before joining King Abdullah University of
 Science and Technology (KAUST), Thuwal, Makkah Province, Saudi
 Arabia as a Professor of Electrical Engineering in 2009. His current
 research interests include the modeling, design, and
 performance analysis of wireless communication systems.
 \end{IEEEbiography}
\end{document}